\documentclass[prd,superscriptaddress,nofootinbib,floatfix,twocolumn]{revtex4}
\pdfoutput=1
\usepackage{bm,amsmath,amssymb,natbib}
\usepackage{ifthen}
\usepackage{aas_macros}
\usepackage{graphicx}
\usepackage[svgnames]{xcolor}
\usepackage{placeins}
\usepackage{hyperref}
\hypersetup{
    pdftitle={Massive Neutrinos and Magnetic Fields in the Early Universe},
    pdfauthor={J. Richard Shaw and Antony Lewis},     % author
    pdfsubject={Cosmology},   % subject of the document
}

\newboolean{showcomments}
\setboolean{showcomments}{true}
\newcommand{\antcomment}[1]{\ifthenelse{\boolean{showcomments}}{{\color{DarkGreen} #1}}{}}
\newcommand{\riccomment}[1]{\ifthenelse{\boolean{showcomments}}{{\color{Maroon} #1}}{}}

\newcommand{\figref}[1]{Fig.~\ref{#1}}

\allowdisplaybreaks[1]

\newcommand{\clh}{\mathcal{H}}
\newcommand{\clr}{\mathcal{R}}
\newcommand{\clf}{\mathcal{F}}

\newcommand\pdiff[2]{\frac{\partial #1}{\partial #2}}
\newcommand\diff[2]{\frac{d #1}{d #2}}

\newcommand\del{\nabla}

\newcommand{\vx}{\mathbf{x}}

\newcommand{\vk}{\mathbf{k}}
\newcommand{\vq}{\mathbf{q}}
\newcommand{\vp}{\mathbf{p}}

\newcommand\vkhat{\hat{\mathbf{k}}}

\newcommand{\khat}{\hat{k}}
\newcommand{\phat}{\hat{p}}
\newcommand{\qhat}{\hat{q}}
\newcommand{\ve}{\mathbf{e}}

\newcommand{\brsc}[1]{{\ensuremath{{\scriptscriptstyle \left(#1\right)}}}}

\usepackage{hyperref}

\begin{document}

\title{Massive Neutrinos and Magnetic Fields in the Early Universe}

\author{J. Richard Shaw}
\email{jrs65@ast.cam.ac.uk}
\author{Antony Lewis}
\homepage{http://cosmologist.info}

\affiliation{Kavli Institute for Cosmology, Madingley Road, Cambridge, CB3 0HA, UK.}
\affiliation{Institute of Astronomy, Madingley Road, Cambridge, CB3 0HA, UK.}

\date{\today}
\begin{abstract}
  Primordial magnetic fields and massive neutrinos can leave an
  interesting signal in the CMB temperature and polarization. We
  perform a systematic analysis of general perturbations in the
  radiation-dominated universe, accounting for any primordial magnetic
  field and including leading-order effects of the neutrino mass. We
  show that massive neutrinos qualitatively change the large-scale
  perturbations sourced by magnetic fields, but that the effect is
  much smaller than previously claimed. We calculate the CMB power
  spectra sourced by inhomogeneous primordial magnetic fields, from
  before and after neutrino decoupling, including scalar, vector and
  tensor modes, and consistently modelling the correlation between the
  density and anisotropic stress sources. In an appendix we present
  general series solutions for the possible regular primordial
  perturbations.
\end{abstract}


\maketitle
\section{Introduction}

The origin of the $10^{-6}$G magnetic fields observed in galaxies and
clusters poses something of a problem for contemporary astrophysics
\cite{Kulsrud:2008}. Recent observations of galaxies at redshift
$z\sim 0.7$---$2$ seem to show the fields were of comparable strength
when the Universe was much younger, disfavouring a large dynamo
amplification from tiny $\sim 10^{-20}$G seed
fields~\cite{Wolfe:2008,Bernet:2008qp}. Tentative observations of
magnetic fields in elliptical galaxies, and a detection in a dwarf
galaxy, also disfavour several dynamo mechanisms because they have
little coherent rotation (see \cite{Widrow:2002} and references
within).  There is also some evidence for $\sim 10^{-8}$G fields
coherent on megaparsec scales~\cite{Lee:2009fw}. It may be possible to
explain these observations in terms of astrophysically generated seed
fields. Another interesting possibility is a primordial seed field.  A
primordial $B \sim 10^{-9}$G (comoving) field could lead to the
observed galactic fields via adiabatic contraction alone, and might
leave an interesting observable signature in the CMB. In this paper we
revisit the calculation of the CMB power spectrum from primordial
inhomogeneous magnetic fields and work towards robust theoretical
predictions that can be used to test the primordial field scenario
with CMB data.  Since primordial magnetic fields are expected to be
exponentially small in most early-universe models, any detection would
be a clear signature of something very interesting.

If a primordial inhomogeneous magnetic field is present it sources
scalar, vector and tensor modes, giving rise to a signal in the CMB
temperature as well as E and B-mode polarization. Previous
calculations have indicated that $\sim 10^{-9}$G fields (comoving) are
detectable \cite{Yamazaki:2006,Paoletti:2008}, but have been
incomplete in several respects. One complication is that the
anisotropic stress due to the magnetic fields becomes compensated by
the neutrino anisotropic stress~\cite{Lewis:2004}, significantly
reducing the perturbations sourced on large scales after neutrino
decoupling.  Recent work by Kojima et al. \cite{Kojima:2008} has
claimed that the presence of massive neutrinos leads to a significant
change in this compensation mechanism, giving rise to a dramatic
enhancement of up to eight orders of magnitude on the large-scale
E-mode polarization power spectrum. For interesting neutrino masses
this would, if true, be a clear signal of primordial magnetic
fields. Clearly this claim merits further investigation, though we
shall ultimately show that the effect, though interesting, is much
smaller than previously claimed.

In the early universe the massive neutrinos are expected to be
relativistic, with the most massive eigenstate only becoming
non-relativistic around recombination or later
\cite{Lesgourgues:2006}. We perform a systematic analysis of the
primordial perturbations to lowest order in the mass, generalizing
previous results for the general primordial perturbation to the
realistic case where one or more of the neutrinos is massive. This
also allows us to calculate the series solutions consistently in the
presence of magnetic fields, and see the leading corrections due to
the neutrino mass effect. We also discuss the tight coupling
approximation, which is useful after the modes come inside the horizon
but before Thomson scattering becomes ineffective.

%[add refs to previous magnetic field calcs]

The calculation of the CMB power spectrum from primordial magnetic
fields is further complicated because the scalar, vector and tensor
sources are all quadratic in the underlying magnetic field, and the
scalar modes have more than one source term. These sources are
correlated, so we show how to calculate the various source power
spectra and correlations from the power spectrum of the magnetic
field, and how to use these for a numerical calculation.

In this work we address the effects of magnetic fields in sourcing
primary anisotropies in the CMB, however, magnetic fields present
after recombination also have an observational effect on the
polarisation by inducing Faraday rotation
\cite{Kosowsky:2005,Scoccola:2004}. Such rotation converts E-mode
polarisation into B-modes with a strong dependence on the frequency of
the radiation $\propto B / \nu^2$ but it is small at the usual
frequencies for CMB observation, and we will neglect it in our
analysis.

Throughout this work we use a 3+1 splitting of General Relativity,
working with a gauge invariant linear perturbation theory similar to
that of Bardeen \cite{Bardeen:1980} and Durrer
\cite{Durrer:1988,Durrer:1994}. Our choice of gauge invariant
variables is chosen as a close analogy to the Conformal Newtonian
Gauge (CNG). We use a metric
\begin{multline}
ds^2 = a(\tau)^2 \Bigl[-(1 + 2A)\, d\tau^2 - 2 B_i\, d\tau\, dx^i \\
  + \left(\delta_{ij} + 2 H_{ij}\right)\, dx^i\, dx^j \Bigr] 
\end{multline}
where we can further decompose $B_i$ and $H_{ij}$ into their scalar,
vector and tensor contributions. Our decompositions are performed in
the same manner as Ref.~\cite{Hu:1997}. In $k$-space the decomposition
for the rank-1 and rank-2 three-tensors are written as (using $B_i$
and $H_{ij}$ as examples)
\begin{align}
B_i & = B Q^\brsc{0}_i + B^\brsc{1} Q^\brsc{1}_i \notag \\
H_{ij} & = H_L \delta_{ij} Q^\brsc{0} + H_T Q^\brsc{0}_{ij} + H^\brsc{1} Q^\brsc{1}_{ij} + H^\brsc{2} Q^\brsc{2}_{ij} 
\end{align}
where the harmonic $Q^\brsc{m}$ functions give the form of each
perturbation type for a specific $k$ mode, with $m = 0$ giving scalar
perturbations, and $m = 1$, $m = 2$, vector and tensor
respectively. In the above it should be understood that we implicitly
sum over the two vector and two tensor modes, for example
\begin{equation}
H^\brsc{1} Q_{ij}^\brsc{1} \equiv H^\brsc{+1} Q_{ij}^\brsc{+1} + H^\brsc{-1} Q_{ij}^\brsc{-1} \, ,
\end{equation}
whilst a quantity like $H^\brsc{1}$ appearing on its own can stand for
either $H^\brsc{+1}$ or $H^\brsc{-1}$ consistent with the context. The
scalar harmonic functions are
\begin{align}
Q^\brsc{0} & = e^{i \vk \cdot \vx} \notag \\
Q_i^\brsc{0} & = - k^{-1} \del_i Q^\brsc{0} = \khat_i e^{i \vk \cdot \vx} \\
Q_{ij}^\brsc{0} & = \left[ k^{-2} \del_i\del_j + \delta_{ij}/3\right] Q^\brsc{0} = -\left(\khat_i\khat_j - \frac{1}{3} \delta_{ij}\right) e^{i \vk \cdot \vx} \; . \notag
\end{align}
The vector harmonics are
\begin{align}
Q_i^\brsc{\pm 1} & = e^\brsc{\pm}_i e^{i \vk \cdot \vx} \notag \\
Q_{ij}^\brsc{\pm 1} & = -\frac{1}{k} \del_{(i} Q^\brsc{\pm 1}_{j)} = i \khat_{(i}e^\brsc{\pm}_{j)}  e^{i \vk \cdot \vx} \, ,
\end{align}
where we decompose our vectors with the helicity basis
\begin{equation}
  e^\brsc{\pm}_i = - \frac{i}{\sqrt{2}} \left( e^{1}_i \pm i e^2_i\right),
\end{equation}
with $\ve^1$ and $\ve^2$ being unit vectors orthogonal to
$\vkhat$. Note that $\ve^\brsc{\pm}\cdot\ve^\brsc{\pm} =
\ve^{\brsc{\pm}*}\cdot\ve^\brsc{\mp} = 0$, whilst
$\ve^{\brsc{\pm}*}\cdot\ve^\brsc{\pm} =
-\ve^\brsc{\pm}\cdot\ve^\brsc{\mp} = 1$. From this we can see that
$Q_{ij}^{\brsc{\pm 1}*} Q^{ij\,\brsc{\pm 1}} = \frac{1}{2}$. For the
tensors we make the further definition of $e_{ij}^\brsc{\pm 2} =
\sqrt{3/2} \; e^\brsc{\pm}_i e^\brsc{\pm}_j$. Using this, the sole
tensor harmonic is
\begin{equation}
Q_{ij}^\brsc{\pm 2} = e^\brsc{\pm 2}_{ij} e^{i \vk \cdot \vx} \; .
\end{equation}
For reference the self contraction of this is $Q_{ij}^{\brsc{\pm 2}*}
Q^{ij\,\brsc{\pm 2}} = \frac{3}{2}$. As we would expect, quantities of
different types are always orthogonal, for example $Q_{ij}^{\brsc{\pm
    2}*} Q^{ij\,\brsc{\pm 1}} = 0$. 

Generally we will drop the superscript on scalar perturbations like
$X^\brsc{0}$, in favour of simply $X$.

As a further illustration, let us examine the perturbations to the
energy-momentum tensor $T_{\mu\nu}$. In the Conformal Newtonian Gauge,
the gauge invariant quantities we use are exactly equivalent to
perturbations of the Energy-Momentum tensor. The density $\Delta$,
velocity $V_i$, pressure $\pi$, and anisotropic stress $\Pi^i_j$
perturbations are defined by
\begin{align}
  T^0_0 &= -\rho (1+ \Delta) \, , \\
  T^0_i &= (\rho + p) V_i \, ,\\
  T^i_j &= p \left[(1 + \pi) \delta^i_j + \Pi^i_j\right] \; ,
\end{align}
where $\rho$ and $p$ are the density and pressure respectively.  As
above we will further decompose the three-vector and tensor quantities
into the different perturbation types.  The velocity three-vector
decomposes as
\begin{equation}
V_i = V Q^\brsc{0}_i + \Omega^\brsc{1} Q^\brsc{1}_i \, ,
\end{equation}
where the vorticity $\Omega^\brsc{1}$ is the vector-type velocity. The
traceless anisotropic stress tensor becomes
\begin{equation}
\Pi^i_j = \Pi Q^\brsc{0}_{ij} + \Pi^\brsc{1} Q^\brsc{1}_{ij} + \Pi^\brsc{2} Q^\brsc{2}_{ij} \; .
\end{equation}
Generally we will rewrite the pressure perturbation $\pi$ in terms of
an entropy type perturbation $\Gamma$ and the density $\Delta$
\begin{equation}
\pi = \Gamma + \frac{c_s^2}{w} \Delta \, .
\end{equation}
Although we will not explicitly demonstrate it, $\Gamma$ is
gauge-invariant. Above we have used $w$ and $c_s^2$, defined as $w = p
/ \rho$ and the sound speed $c_s^2 = \dot{p} / \dot{\rho}$.

We will restrict ourselves to a flat geometry throughout this work.

%%% Local Variables:
%%% mode: latex
%%% TeX-master: "paper"
%%% End:

\section{Neutrino Perturbations}

\subsection{Kinetic Theory}

To describe the behaviour of neutrinos in the early universe, we must
turn to the full machinery of the Boltzmann equation. We start with
the phase space distribution of the particle density on a spatial
hypersurface, defined by
\begin{equation}
dN = f_\nu(x^i,P_j,\tau)\, d^3x\, d^3P
\end{equation}
where $P_i$ is the canonical 3-momentum, the spatial part of the
covariant 4-momentum $P_\mu$. The primary quantity we will require is
the energy-momentum tensor which is determined from the distribution
function $f_\nu$ by
\begin{equation}
\label{eq:tmn_int}
T^\mu_{\phantom{\mu}\nu} = \int \frac{d^3P\, (-g)^{1/2}}{P_0} P^\mu P_\nu f_\nu (x^i,P_j,\tau) \; .
\end{equation}

Following the convention in the literature \cite{Durrer:1994,Ma:1995}
we will re-express the distribution function in terms of quantities in
the frame of a comoving observer. We use the locally Minkowski tetrad
$e_a^\mu$ satisfying $g_{\mu\nu} e_a^\mu e_b^\nu = \eta_{ab}$. In terms
of the co-ordinate basis, and where we have avoided fixing a gauge
\begin{subequations}
\begin{align}
e_0 & = a^{-1}\left[(1 - A)\, \partial_0 - B^i \,\partial_i \right] \; ,\notag\\
e_i & = a^{-1}\left[(1 - H_L)\, \partial_i - H_i^j \, \partial_j \right] \; ,
\end{align}
\end{subequations}
with $H_i^j$ containing the trace-free scalar, vector and tensor
contributions. This allows us to write the momentum in terms of
quantities measured in the comoving tetrad $P = P^\mu \partial_\mu =
\pi^a e_a$, where $\pi^0$ is the observed energy and $\pi^i$ the
momentum in that frame. Applying Hamilton's equations to the system
implies that the conjugate momenta will remain constant in a purely
FRW universe. This means the proper 4-momenta will decay away with
$a^{-1}$. In order to remove this redshifting of the energy and
momenta we will write them in terms of the scaled quantities
$\epsilon$ and $q$ defined by
\begin{align}
\pi^0 & = \epsilon / a \; , \notag \\
\pi^i & = q n^i / a \; ,
\end{align}
where $n^i$ is the unit vector in the direction of the momentum. Both
$\epsilon$ and $q$ are constant on the background by definition. By
considering $P\cdot P = \pi_a \pi^a$ we find a slightly modified
energy momentum relation
\begin{equation}
\epsilon(q) = (q^2 + a^2 m^2)^{1/2} \; .
\end{equation}

Prior to their decoupling, neutrinos are in approximate thermal
equilibrium with the rest of the Universe. Considering only the
unperturbed case for the moment, the phase-space distribution function
of the neutrinos $f_{\nu 0}$ will be Fermi-Dirac at a universal
temperature. We expect this distribution to be isotropic and
homogenous, and thus only be a function of the momentum magnitude $q$
(in the guise of the comoving energy) and the time $\tau$ (by virtue
of the temperature). Therefore it takes the form
\begin{equation}
f_{\nu 0}(q, \tau) = \frac{g_s}{h_p^3}\frac{1}{e^{E(q)/k_B T(\tau)} + 1} \; ,
\end{equation}
where the neutrino energy measured by a comoving observer is $E =
\epsilon / a$. As the temperature decreases with $1/a$, the
combination $E(q) / k_B T = \epsilon(q) / k_B T_0$ is constant,
depending on $T_0$, the temperature today.

At neutrino decoupling, this distribution becomes frozen in. 
%This occurs much earlier than any epoch of interest: 
The neutrino mass is insignificant compared to any thermal energy, so
its contribution can be neglected in the distribution function. This
allows us to set $\epsilon = q$ (within the distribution only),
leaving the unperturbed function as
\begin{equation}
f_{\nu 0}(q) = \frac{g_s}{h_p^3}\frac{1}{e^{q/k_B T_0} + 1} \; .
\end{equation}

We will define the first order perturbations to the distribution
$\psi_\nu$ by
\begin{equation}
f_\nu(x^i,P_j,\tau) = f_{\nu 0}(q)\left[1 + \psi_\nu(x^i,q,n_j,\tau)\right] \; ,
\end{equation}
with $\psi_\nu$ containing the scalar, vector and tensor
contributions. This quantity is gauge dependent; later we will form a
gauge invariant equivalent.

We want to rewrite the integral \eqref{eq:tmn_int} in terms of our
comoving quantities, retaining terms up to first order. Firstly, the
term $d^3P\, (-g)^{1/2} / P_0$ forms a co-ordinate invariant measure
for the integration, and can be re-written in terms of the comoving
quantities
\begin{equation}
\frac{d^3P\, (-g)^{1/2}}{P_0} = a^{-2} dq d\Omega_n \frac{q^2}{\epsilon} \; .
\end{equation}
This removes the metric perturbations contained within the integration
measure. Re-expressing the $P^\mu P_\nu$ generates a plethora of
terms, including terms first-order in the metric perturbations. However
these terms all depend upon a single power of the momentum direction
$n_i$ and couple only with the isotropic distribution $f_{\nu 0}$;
they are thus eliminated by their symmetry. The remaining terms are
simply
\begin{equation}
P^\mu P_\nu = a^{-2} (\epsilon \delta^\mu_0 + q n^i \delta^\mu_i) ( -\epsilon \delta_\nu^0 + q n_i \delta^i_\nu ) + \ldots
\end{equation}
Decomposing into distinct components this leaves us with
\begin{align}
T^0_{\phantom{0}0} &=  - a^{-4} \int q^2 dq d\Omega_n\, \epsilon\, f_{\nu 0}(q) \left[1 + \psi_\nu\right] \; , \notag \\
T^0_{\phantom{0}i} &= a^{-4} \int q^2 dq d\Omega_n\, q n_i\, f_{\nu 0} \psi_\nu \; , \\
T^i_{\phantom{i}j} &= a^{-4} \int q^ 2dq d\Omega_n\, \frac{q^2}{\epsilon} n^i n_j \, f_{\nu 0}(q) \left[1 + \psi_\nu\right] \notag \; ,
\end{align}
valid in all gauges up to first order.

The evolution of the distribution function is governed by the
collisionless Boltzmann (or Vlasov) equation, which simply expresses
that, without collisions, the number of particles is conserved along a
trajectory in phase space
\begin{equation}
\label{eq:boltzmann_def}
\frac{D f}{D \tau} = \pdiff{f}{\tau} + \diff{x^i}{\tau} \pdiff{f}{x^i} + \diff{q}{\tau} \pdiff{f}{q} + \diff{n_i}{\tau}\pdiff{f}{n_i} = 0 \; .
\end{equation}
In order to address the evolution of the perturbations to the
distribution function, we need to separate this out into equations for
the background, and the separate perturbation types. We address this
in subsequent sections.

\subsection{Background Quantities}

At zeroth order the collisionless Boltzmann equation simply shows that
the distribution remains constant, that is, $f_{\nu 0}$ is independent
of time. At this order the energy-momentum tensor can be described
fully in terms of density and the pressure. The density is given by
\begin{equation}
\rho_\nu = 4\pi a^{-4} \int q^2 dq \epsilon f_{\nu 0}(q) \; .
\end{equation}
With a non-zero neutrino mass the pressure is no longer simply related
to the density. It's instead
\begin{equation}
p_\nu = \frac{4\pi}{3} a^{-4} \int q^2 dq \frac{q^2 }{\epsilon}f_{\nu 0}(q) \; .
\end{equation}
As we would expect, the equation of state $w_\nu = p_\nu / \rho_\nu$
is still defined by the ratio of these two quantities, yielding a mass
dependent $w \ne 1/3$.

\subsection{Scalar Perturbations}

Initially we will just address scalar perturbations to the
distribution function $\psi_\nu^{\brsc{0}}$, considering vectors and
tensors later on.

We will perform a harmonic expansion of all our quantities. In
flat-space this is just the Fourier transform. The Boltzmann equation
\eqref{eq:boltzmann_def} can be expanded at first order including all
the (non-gauge fixed) metric perturbations \cite{Durrer:1988} giving
\begin{multline}
\dot{\psi}_\nu^{\brsc{0}} + i k \mu \frac{q}{\epsilon} \psi_\nu^{\brsc{0}} = \frac{d \ln{f_{\nu 0}}}{d\ln{q}} \\
\times\left[i k \mu \frac{\epsilon}{q} A + B k \mu^2 + \dot{H}_L - \left(\mu^2 - 1/3 \right) \dot{H}_T \right]\; ,
\end{multline}
where $\mu = n^i k_i$ and the dot is the derivative with respect to
conformal time $\tau$. To move to a gauge-invariant formalism we
follow Durrer and Straumann \cite{Durrer:1988} and define a new
gauge-invariant distribution perturbation
\begin{equation}
\Psi_\nu^{\brsc{0}}= \psi_\nu^{\brsc{0}} - \sigma \frac{d \ln{f_{\nu 0}}}{d \ln{q}} \left[\frac{\clh }{k} + i \frac{\epsilon}{q} \mu \right] \; ,
\end{equation}
where $\sigma$ is the shear on spatial hypersurfaces $\sigma =
\dot{H_T} / k - B$, and $\clh = \dot{a}/a$ is the conformal Hubble
parameter. This differs from Durrer's definition in that we have
chosen our definition to coincide with the CNG result 
%to keep ananalogy with the Conformal Newtonian Gauge, with the
(added terms vanish in a zero-shear gauge). 
For comparison Durrer's
invariant perturbation $\clf$ and our definition are linked via
\begin{equation}
\clf^{\brsc{0}} = f_{\nu 0} \left(\Psi_\nu^{\brsc{0}} - \frac{d \ln{f_{\nu 0}}}{d \ln{q}} \Phi \right) \; .
\end{equation}
Instead of the scalar metric perturbations we will use the
gauge-invariant Bardeen potentials
\begin{align}
\Psi & = A + \frac{1}{k}\left(\dot{B} + \clh B\right) - \frac{1}{k^2} \left(\ddot{H}_T + \clh \dot{H}_T\right) \notag \\
 & = A - \clh \sigma / k - \dot{\sigma} / k \\
\Phi & = -H_L - \frac{1}{3}H_T - \frac{\clh}{k} B + \frac{\clh}{k^2} \dot{H}_T \\
& = -\clr + \clh \sigma / k \; .\notag
\end{align}
The potential $\Psi$ should not be confused with the distribution
perturbation $\Psi_\nu^\brsc{m}$. Usually the context will make this
clear. In the above, $\clr$ is the 3-Ricci scalar
\begin{equation}
\clr = H_L + \frac{1}{3}H_T \; .
\end{equation}
Written in terms of gauge-invariant quantities, the Boltzmann equation becomes
\begin{equation}
\label{eq:bl_scalar}
\dot{\Psi}_\nu^{\brsc{0}} + i k \mu \frac{q}{\epsilon} \Psi_\nu^{\brsc{0}} + \frac{d \ln{f_{\nu 0}}}{d\ln{q}} \left[\dot{\Phi} - i k \mu \frac{\epsilon}{q} \Psi \right] = 0 \; .
\end{equation}
As mentioned, our choice of gauge invariant variables is designed such
that this is equivalent to the CNG version.

The dependence on the momentum direction within the Boltzmann equation
makes a direct solution tricky. We take the standard approach and
expand out into an angular basis. Whilst for scalar perturbations it
suffices to expand in the Legendre polynomials $P_l(\mu)$, for vector
and tensor perturbations it is much more convenient to use a method
similar to Ref.~\cite{Hu:1997}, where we expand out into spherical
harmonics $Y_l^m(\theta,\phi)$. Under this expansion, the different
types of perturbations are separated by their $m$ value, with scalar
($m = 0$), vector ($m = 1$) and tensor ($m = 2$) modes all evolving
separately in the usual manner. Expanding out the entire distribution
perturbation
\begin{equation}
\Psi_\nu= \sum^\infty_{l = 0}\sum_{m = 0}^l (-i)^l \sqrt{\frac{4\pi}{2 l + 1}} \Psi_{\nu l}^{\brsc{m}}(k_i, q) Y_l^m(n^j) \; ,
\end{equation}
we can relate the momentum integrals of the multipole moments to the
standard gauge invariant perturbations
\begin{align}
\Delta_\nu^\brsc{0}(k_i) & = \frac{4\pi}{\rho_\nu a^4} \int q^2 dq \epsilon f_{\nu 0}(q) \Psi_{\nu 0}^{\brsc{0}}(k_i, q) \; , \notag \\
V_\nu^\brsc{0}(k_i) & = \frac{4 \pi}{3 (\rho_\nu + p_\nu) a^4} \int q^2 dq q f_{\nu 0}(q) \Psi_{\nu 1}^{\brsc{0}}(k_i, q) \; , \\
\Pi_\nu^\brsc{0}(k_i) & = \frac{4 \pi}{5 p_\nu a^4} \int q^2 dq \frac{q^2}{\epsilon} f_{\nu 0}(q) \Psi_{\nu 2}^{\brsc{0}}(k_i, q) \; . \notag
\end{align}

\subsection{Thermal Perturbations}

The most natural perturbation that could be set up is from a purely
thermal distribution, where we perturb the neutrinos by having a
position and direction dependent change to the temperature. To take
this into account let us re-write our distribution perturbation in a
slightly different manner. The total distribution
\begin{align}
f_\nu(x^i,q, n_j,\tau) & = \frac{g_s}{h_p^3}\frac{1}{e^{q/k_B T_0 \left(1+ \theta_\nu\right)} + 1} \\
& =f_{\nu 0}(q)\left[1 - \frac{d\ln{f_{\nu 0}}}{d\ln{q}} \theta_\nu(x^i,q, n_j, \tau)\right] \notag \; .
\end{align}
For generality we leave $\theta$ a function of $q$. For a pure
temperature perturbation it must be temperature independent. Relating
this to our previous gauge invariant perturbation $\Psi_\nu$ we find
that
\begin{equation}
\Psi_\nu= - \frac{d \ln{f_{\nu 0}}}{d \ln{q}} \left[\theta_\nu(x^i,q,n_j,\tau) + \sigma \left(\frac{\clh }{k} + i \frac{\epsilon}{q} \mu\right) \right] \, ,
\end{equation}
and from this construct a temperature-like gauge-invariant
perturbation
\begin{equation}
\Theta_\nu(x^i,q,n_j,\tau) = \theta_\nu + \sigma \left(\frac{\clh }{k} + i \frac{\epsilon}{q} \mu\right) \; ,
\end{equation}
such that $\Psi_\nu= - \frac{d \ln{f_{\nu 0}}}{d \ln{q}}
\Theta_\nu$. Substitution into the Boltzmann equation
\eqref{eq:bl_scalar} produces
\begin{equation}
\dot{\Theta}_\nu + i k \mu \frac{q}{\epsilon} \Theta_\nu - \left[\dot{\Phi} - i k \mu \frac{\epsilon}{q} \Psi \right] = 0 \; .
\end{equation}
When $m = 0$ or $\tau \rightarrow 0$, $\epsilon = q$ and the Boltzmann
equation becomes momentum independent; the perturbation remains purely
thermal. However even perturbations which start in a purely thermal
state evolve away from it as the mass becomes important. For this
reason we must keep $\Theta_\nu$ a function of $q$, though we will
restrict ourselves to purely thermal initial conditions.

\section{Mass Expansion}
\label{sec:massexpansion}
To treat massive neutrinos in the early Universe, when the mass is
small in comparison to the typical momentum (approximately $k_B
T_\nu$), we will expand the system to first order in the neutrino mass
squared. This will allow us to directly tackle the integrated
distribution function, making it possible to find initial conditions
up to this order in the neutrino mass. For a more general approach see
\cite{Lewis:2002}.

In both the integrals for the energy-momentum tensor and the Boltzmann
equation itself the mass dependence comes in from factors of $\epsilon
/ q$ or its inverse. Expanding this out gives $\epsilon / q = 1 + m^2
/ 2q^2 + \dotsb$ (with a minus for the inverse). For the background
quantities performing this expansion gives
\begin{align}
\rho_\nu & = 4\pi a^{-4} \int q^2 dq \epsilon f_{\nu 0}(q) \notag \\
 & = 4\pi a^{-4} \int q^2 dq \, q \, f_{\nu 0}(q) \left[1 + \frac{1}{2}\frac{m^2 a^2}{q^2} + \ldots\right] \\
 & = \rho_{\nu 0} \left( 1 + \frac{1}{2} \bar{m}^2 a^2 \right) + \ldots \; ,\notag
\end{align}
where $\rho_{\nu 0}$ is the density for massless neutrinos, and the scaled
mass $\bar{m}^2 = m^2 / \bar{q}^2$, with the $\bar{q}^2$ factor being defined via
\begin{equation}
\frac{1}{\bar{q}^2} = \frac{\int q^2 dq \, q \, q^{-2} f_{\nu 0}(q)}{\int q^2 dq \, q \, f_{\nu 0}(q)} \; ,
\end{equation}
which is essentially the momentum averaged inverse square
momentum. This depends only on the background distribution, and is
time independent. Thus we factor it into the new dimensionless mass
$\bar{m}$. In terms of background quantities
\begin{equation}
\bar{m} = \sqrt{\frac{10}{7 \pi^2}} \frac{m}{k_B T_0} \; .
\end{equation}

As with the density we expand up to $m^2$ for the background
pressure $p_\nu$ and the equation of state $w_\nu$ giving the
form
\begin{align}
p_\nu & = p_{\nu 0} \left( 1 - \frac{1}{2} \bar{m}^2 a^2 \right ) \; , \notag \\
w_\nu & = p_\nu / \rho_\nu  = \frac{1}{3} \left( 1 - \bar{m}^2 a^2 \right) \; .
\end{align}

The perturbed quantities are slightly more tricky. For example taking
$\Delta_\nu$ and expanding out in mass gives,
\begin{equation}
  \Delta_\nu = \frac{4\pi}{\rho_\nu a^4} \int q^2 dq \, q \, f_{\nu 0}(q) \, \Psi_{\nu 0}  \left(1 + \frac{1}{2}\frac{m^2 a^2}{q^2}\right) \, ,
\end{equation}
with a similar pattern for the other perturbations. Schematically we
have two types of terms we need to integrate; $\int q^2 dq \, q \,
f_{\nu 0} \Psi_\nu$ and $\int q dq f_{\nu 0} \Psi_\nu$. For a moment
let us consider what happens to these integrals for thermal
perturbations in the case of massless neutrinos
\begin{align}
\int q^2 dq \, q \, f_{\nu 0}\Psi_\nu &= -\Theta_\nu \int q^2 dq \, q \, f_{\nu 0} \diff{\ln{f}}{\ln{q}} \notag \\
& = 4 \Theta_\nu \left( \frac{a^4 \rho_{\nu 0}}{4 \pi}\right) \; ,
\end{align}
and similarly the second integral equates to
\begin{equation}
\int q dq f_{\nu 0}\Psi_\nu = 2 \Theta_\nu \left( \frac{a^4 \rho_{\nu 0}}{4 \pi}\right) \frac{1}{\bar{q}^2} \; .
\end{equation}
From this we can make the connection that, at zeroth order in the mass
expansion, the second integral is linked to the first via
\begin{equation}
\label{eq:int_equiv}
\int q dq f_{\nu 0}\Psi_\nu = \frac{1}{2\bar{q}^2} \int q^2 dq \, q \, f_{\nu 0}\Psi_\nu + O(\bar{m}^2) \; ,
\end{equation}
and as the second integral appears at first order in $m^2$ in the mass
expansion, we can use this relation to simplify the expression for
$\Delta_\nu$ above, giving
\begin{equation}
  \Delta_\nu = \frac{4\pi}{\rho_\nu a^4} \left(1 + \frac{1}{4}\bar{m}^2 a^2 \right) \,\int q^2 dq \, q \, f_{\nu 0} \, \Psi_{\nu 0} \, .
\end{equation}
This happens similarly with the other perturbed quantities, and allows
us to follow the convention of forming a momentum integrated function
\begin{align}
\label{eq:fdef}
F(k_i,\mu, \tau) & = \frac{\int q^2 dq \, q \, f_{\nu 0}(q) \Psi_{\nu}(k_i,q,\mu)}{\int q^2 dq \, q \, f_{\nu 0}(q)} \notag \\
&= \frac{4 \pi a^{-4}}{\rho_{\nu 0}} \int q^2 dq \, q \, f_{\nu 0}(q) \Psi_{\nu}(k_i,q,\mu) \; .
\end{align}
As with the distribution perturbation, we will expand $F$ into
spherical harmonics
\begin{equation}
F(k_i, \mu, \tau) = \sum^\infty_{l = 0} (-i)^l \sqrt{\frac{4\pi}{2 l + 1}} F_l^{\brsc{0}}(k_i, \tau) Y_l^0(\mu) \; .
\end{equation}
Each moment $F_l$ takes the same form as the integrated $F$ of
\eqref{eq:fdef} with the $\Psi_\nu$ being replaced by $\Psi_{\nu l}$.

These lead to a succinct form for the perturbations
\begin{align}
\Delta_\nu & = \frac{\rho_{\nu 0} F_{0} \left(1 + \frac{1}{4}\bar{m}^2 a^2\right)}{\rho_0 \left(1 + \frac{1}{2}\bar{m}^2 a^2\right)} \notag \\
& = F_0 \left(1 - \frac{1}{4}\bar{m}^2 a^2 \right)
\end{align}
to order $m^2$. Similarly,
\begin{align}
V_\nu & = \frac{1}{4} F_1 \left(1 - \frac{1}{4}\bar{m}^2 a^2\right) \; , \notag \\
\Pi_\nu & = \frac{3}{5} F_2 \left(1 + \frac{1}{4}\bar{m}^2 a^2 \right) \; .
\end{align}
All that we need to do to have a working prescription for calculating
the massive neutrino evolution is to turn the Boltzmann equation into
a hierarchy for solving for the $F_l^{\brsc{m}}$. First we take the
Boltzmann equation and expand to first order in $m^2$, then
integrate it over $\int q^2 dq \, q \, f_{\nu 0}(q)$ and divide by the same
quantity to produce an equation for the evolution of $F$. We employ
the same trick as above (in Eq.~\eqref{eq:int_equiv}) to turn the
$m^2 / q^2$ quantities into terms in $F$. This results in
\begin{multline}
\dot{F} + i k \mu F \left(1 - \frac{1}{4}\bar{m}^2 a^2 \right) \\ = 4 \dot{\Phi} - 4 i k \mu \Psi \left(1 + \frac{1}{4}\bar{m}^2 a^2 \right) \; .
\end{multline}
We then substitute a spherical harmonic expansion for $F$, and using
the identity that
\begin{equation}
\label{eq:mu_y}
\mu Y_l^{m} = \sqrt{\frac{l^2 - m^2}{4l^2 - 1}} Y_{l-1}^{m} + \sqrt{\frac{(l+1)^2 - m^2}{4(l+1)^2 - 1}} Y_{l+1}^{m}
\end{equation}
we obtain the hierarchies for $F$. Separating these out into coupled
equations for each $l$ gives three distinct cases. For the monopole
($l = 0$)
\begin{subequations}
\label{eq:neutrino_eq}
\begin{equation}
\dot{F}_0 + \frac{k}{3} F_1 \left(1 - \frac{1}{4}\bar{m}^2 a^2 \right) = 4 \dot{\Phi} \; .
\end{equation}
For the dipole ($l = 1$)
\begin{equation}
\dot{F}_1 + \frac{k}{5} \left(1 - \frac{1}{4}\bar{m}^2 a^2\right) \left[ 2 F_2  - 5 F_0 \right] = (4 + \bar{m}^2 a^2) k \Psi \; .
\end{equation}
Finally for the quadrupole and higher moments ($l \geq 2$)
\begin{equation}
\dot{F}_l + k \left(1 - \frac{1}{4}\bar{m}^2 a^2\right) \left[ \frac{l + 1}{2 l + 3} F_{l+1} - \frac{l}{2 l - 1} F_{l-1} \right] = 0 \; .
\end{equation}
\end{subequations}
As should be expected sending $\bar{m} \rightarrow 0$ takes everything
to the well known massless case.

Taking this mass expansion to higher order becomes more difficult:
Taylor expanding the background quantities inside the integral
produces divergent integrals at order $m^4$ and above.  We show how to
do the higher-order expansion in Appendix~\ref{app:higherorder}; this
shows that the leading-order mass expansions are correct to ${\cal
  O}(m^4\log(m))$.
%\antcomment{sort out $m$ vs $m_n$}

\section{Vector Perturbations}

To consider vector perturbations we proceed down a similar line to the
scalar perturbations. Though not entirely free of gauge issues, many
of the complexities will disappear. Firstly, whilst there are two
vector-type metric perturbations $B^{\brsc{1}}$ and $H^{\brsc{1}}$, we
have one degree of gauge freedom, and so only one perturbation can be
relevant. As before, rather than fixing a gauge we form one
gauge-invariant variable. For vector perturbations, the shear-like
perturbation $\sigma^{\brsc{1}} = \dot{H}^{\brsc{1}} / k -
B^{\brsc{1}}$ is gauge-invariant and we use this as our metric
variable.
%\antcomment{the shear itself isn't gauge invariant; you mean the shear in the 
%zero-vorticity frame? (which I suppose is implicit in being able to define a globally orthogonal hypersurface, but might be clearer)}

The vector contribution to the distribution function
$\Psi_\nu^{\brsc{1}}$ is itself gauge-invariant, (see
\cite{Durrer:1994}), and thus the Boltzmann equation governing it is
\begin{equation}
\dot{\Psi}_\nu^{\brsc{1}} + i k \mu \frac{q}{\epsilon} \Psi_\nu^{\brsc{1}} - \frac{d \ln{f_{\nu 0}}}{d\ln{q}} n^i n^j \khat_i e_{j}^\brsc{1} k \sigma^{\brsc{1}} = 0 \; .
\end{equation}
There are two contributions to the energy-momentum tensor: the
velocity, $v_\nu^{\brsc{1}}$, which is gauge dependent, and the
anisotropic stress $\Pi_\nu^{\brsc{1}}$ which is gauge-invariant. As
a gauge-invariant velocity, we use the neutrino vorticity
$\Omega^{\brsc{1}}_\nu = v_\nu^{\brsc{1}} - B^{\brsc{1}}$ which is
conveniently related to the distribution perturbation. The two
neutrino perturbations are therefore
\begin{align}
\Omega_\nu^{\brsc{1}}(k_i) & = \frac{4 \pi}{3 (\rho_\nu + p_\nu) a^4} \int q^2 dq q f_{\nu 0}(q) \Psi_{\nu 1}^{\brsc{1}}(k_i, q) \; , \notag \\
\Pi_\nu^{\brsc{1}}(k_i) & = \frac{8\sqrt{3}\, \pi}{15 p_\nu a^4} \int q^2 dq \frac{q^2}{\epsilon} f_{\nu 0}(q) \Psi_{\nu 2}^{\brsc{1}}(k_i, q) \; .
\end{align}
Performing the same momentum integration as for the scalars (with the
same restriction to thermal modes), we find an equation governing the
momentum-integrated $F^\brsc{1}$
\begin{equation}
\dot{F}^{\brsc{1}} + i k \mu F^{\brsc{1}} \left(1 - \frac{1}{4}\bar{m}^2 a^2 \right)= - 4 \sqrt{\frac{4\pi}{15}} Y_2^1 k \sigma^{\brsc{1}} \; ,
\end{equation}
where we have used that $i n^i n^j \khat_i e_{j}^\brsc{1} =
\sqrt{4\pi/15} Y_2^1$. Inserting the spherical harmonic expansion, and
using the identity \eqref{eq:mu_y}, the moments of the Boltzmann
equation are for $l = 1$
\begin{equation}
\dot{F}_1^{\brsc{1}} + \frac{k\sqrt{3}}{5} F_2^{\brsc{1}} \left(1 - \frac{1}{4}\bar{m}^2 a^2\right) = 0 \; ,
\end{equation}
for $l = 2$
\begin{equation}
\dot{F}_2^{\brsc{1}} + k  \left(1 - \frac{1}{4}\bar{m}^2 a^2\right) \left[\frac{\sqrt{8}}{7} F_3^{\brsc{1}} - \frac{\sqrt{3}}{3} F_2^\brsc{1}\right] = \frac{4}{\sqrt{3}} k \sigma^{\brsc{1}}
\end{equation}
and for $l > 2$
\begin{multline}
\dot{F}_l^{\brsc{1}} + k \left(1 - \frac{1}{4}\bar{m}^2 a^2\right) \\ \times\left[ \frac{\sqrt{(l+1)^2-1} }{2 l + 3} F_{l+1}^{\brsc{1}} - \frac{\sqrt{l^2-1}}{2 l - 1} F_{l-1}^{\brsc{1}}\right] = 0 \; .
\end{multline}
Finally we need to rewrite both $\Omega_\nu^\brsc{1}$ and
$\Pi_\nu^\brsc{1}$ in terms of the integrated $F^\brsc{1}_l$
functions. These are nearly identical to the scalar equivalents, only
with different coefficients
\begin{align}
\Omega^\brsc{1}_\nu(k_i) & = \frac{1}{4} F^\brsc{1}_1 \left(1 - \frac{1}{4}\bar{m}^2 a^2\right) \; , \notag \\
\Pi^\brsc{1}_\nu(k_i) & = \frac{2\sqrt{3}}{5} F^\brsc{1}_2 \left(1 + \frac{1}{4}\bar{m}^2 a^2 \right) \; .
\end{align}

\section{Tensor Modes}

As is well known, tensor perturbations are manifestly gauge invariant
and so we need not concern ourselves with any gauge issues. Other than
that we follow the same track as for the scalar perturbations. The
Boltzmann equation for tensor modes takes on the form
\begin{equation}
\dot{\Psi}_\nu^{\brsc{2}} + i k \mu \frac{q}{\epsilon} \Psi_\nu^{\brsc{2}} - \frac{d \ln{f_{\nu 0}}}{d\ln{q}} n^i n^j \dot{H}_{ij}^{\brsc{2}} = 0 \; .
\end{equation}
We then momentum integrate the equation to produce a single equation
in terms of the $F^{\brsc{2}}$. Again we have restricted ourselves to
initially thermal perturbations, giving
\begin{equation}
\label{eq:t_boltzmann}
\dot{F}^{\brsc{2}} + i k \mu F^{\brsc{2}} \left(1 - \frac{1}{4}\bar{m}^2 a^2 \right)= -4 n^i n^j e^\brsc{\pm 2}_{ij} \dot{H}^{\brsc{2}} \; .
\end{equation}
Using the helicity basis, the quantity $n^i n^j e_{ij}^{\brsc{\pm 2}}$
is simply written in terms of spherical harmonics as
\begin{equation}
\label{eq:H_decomp}
n^i n^j e_{ij}^{\brsc{2}} = -\sqrt{\frac{4\pi}{5}} Y^2_2 \; .
\end{equation}
To finish off, we need to rewrite the Boltzmann equation
\eqref{eq:t_boltzmann} with a spherical harmonic decomposition. There
are no relevant contributions from the $l = 0$, $l = 1$ and $m \ne \pm
2$ terms in the sum. For $l = 2$ we have
\begin{equation}
\dot{F}_2^{\brsc{2}} + \frac{k \sqrt{5}}{7} F_3^{\brsc{2}} \left(1 - \frac{1}{4}\bar{m}^2 a^2\right) = 4 \dot{H}^{\brsc{2}},
\end{equation}
and for $l > 2$ we require
\begin{multline}
\dot{F}_l^{\brsc{2}} + k \left(1 - \frac{1}{4}\bar{m}^2 a^2\right) \\ \times\left[ \frac{\sqrt{(l+1)^2-4} }{2 l + 3} F_{l+1}^{\brsc{2}} - \frac{\sqrt{l^2-4}}{2 l - 1} F_{l-1}^{\brsc{2}}\right] = 0 \; .
\end{multline}
The only tensor contribution to the energy momentum tensor comes from
the anisotropic stress $\Pi_\nu$. This is easily expressed in terms of the
expanded $F^{\brsc{2}}$ with
\begin{equation}
\Pi_\nu^{\brsc{2}} = \frac{2}{5} \left(1 + \frac{1}{4} \bar{m}^2 a^2\right) F_2^{\brsc{2}} \; .
\end{equation}

%%% Local Variables:
%%% mode: latex
%%% TeX-master: "paper"
%%% End:

\section{Evolution Equations}

The behaviour of the early Universe is accurately described by linear
perturbation theory, reducing to a system of coupled linear
differential equations. We have discussed the perturbation equations
for the neutrinos in previous sections. Here we briefly describe the
remaining equations for the evolution of the metric potentials and the
other matter species.

The $3+1$ splitting of the Einstein equation $G_{\mu\nu} = 8\pi G
T_{\mu\nu}$ decomposes into sets of equations for each of the scalar,
vector and tensor contributions. For the scalar perturbations we have
four equations generated by the splittings. There are two equations
formed by the $(00)$ and $(0i)$ components
\begin{subequations}
\label{eq:einstein}
\begin{align}
k^2 \Phi & = -\frac{3}{2}\clh^2\left[\Delta + 3(1 + w)\frac{\clh}{k} V\right]\; , \\
k(\dot{\Phi} + \clh \Psi) & =\frac{3}{2} \clh^2\left(1+w\right) V  \; ,
\end{align}
where the first is the equivalent of the classical Poisson
equation. The spatial part ($ij$) splits into two further equations
from the trace and traceless parts. The equation from the trace is
\begin{multline}
\ddot{\Phi}+\clh (\dot{\Psi} + 2 \dot{\Phi})+\left(2\dot{\clh} + \clh^2\right)\Psi+\frac{1}{3}k^2(\Phi - \Psi) \\=\frac{3}{2} \clh^2\left(c_s^2\Delta + w \Gamma\right) ,
\end{multline}
where $\Gamma$ is the perturbation to the entropy of the system, and
$c_s^2 = \dot{p}/\dot{\rho}$ is the total sound speed of all the
matter species. The final equation is from the traceless part
\begin{equation}
\label{eq:dyn2}
k^2(\Phi - \Psi) =3 \clh^2 w \Pi \, .
\end{equation}
\end{subequations}
There are two vector equations, one from the $(0i)$ part, and the
second from the vector contribution to the $(ij)$ components:
\begin{subequations}
\begin{align}
  k^2 \sigma^\brsc{1} & = - 6 \clh^2 \left(1 + w\right) \Omega^\brsc{1} \; ,\\
  k \left(\dot{\sigma}^\brsc{1} + 2\clh \sigma^\brsc{1}\right) & = 3 \clh^2 w \Pi^\brsc{2} \; .
\end{align}
\end{subequations}
There is a single equation for the tensor modes
\begin{equation}
\ddot{H}^\brsc{2} + 2 \clh \dot{H}^\brsc{2} + k^2 H^\brsc{2} = 3 \clh^2 w \Pi^\brsc{2}\; .
\end{equation}

The matter evolution equations are well known and are most generally
derived from the Boltzmann equation, here we will just give the
results. We consider the standard three matter species beyond
neutrinos: baryons, photons and cold dark matter, giving their
perturbations in terms of $\Delta$, $V$ and $\Pi$ as before.

The matter species have essentially no velocity dispersion, the
anisotropic stress and higher momentum moments are all zero. Hence
they contribute only to scalar and vector modes and can be described
entirely in terms of $\Delta$, $V$ and $\Omega$. Simplest is dark
matter as it has no interactions. For the scalars
\begin{subequations}
\begin{align}
\dot{\Delta}_c & =-k\, V_c + 3 \dot{\Phi} \; ,\\
\dot{V}_c & = -\clh\, V_c + k \, \Psi \; ,\\
\intertext{and for the vectors}
\dot{\Omega}_c^\brsc{1} & = -\clh \Omega_c^\brsc{1} \; .
\end{align}
\end{subequations}
We can see that any vector solution for CDM must be decaying, and so
we will neglect it.

The baryons couple to the photons via Thomson scattering, but also
interact with any magnetic field via the Lorentz force giving an extra
source term
%\antcomment{have magnetic quantities been defined at this point? Also $c_s^2$ here is inconsistent with previous definition in terms of the background}
\begin{subequations}
\begin{align}
\dot{\Delta}_b & =-k\, V_b + 3 \dot{\Phi} \; ,\\
\dot{V}_b & = -\clh\, V_b + k \, c_{s,b}^2 \Delta_b + k \, \Psi  + R \tau_c^{-1} \left(V_\gamma - V_b\right) \notag \\
& \hspace*{20pt}+ \frac{1}{2} k R \left(\frac{1}{2} \Delta_B -  w_\gamma \Pi_B^\brsc{0}\right) \; ,
\end{align}
where the baryon sound speed is $c_{s,b}^2 = \delta p_b/
\delta\rho_b$. The two quantities $\Delta_B$ and $\Pi_B$ are the
magnetic equivalents of the density and anisotropic stress
perturbations. We make a thorough definition in the next section. The
vector equation is
\begin{equation}
\dot{\Omega}_b^\brsc{1} = -\clh \Omega_b^\brsc{1} + R \tau_b^{-1} \left(\Omega^\brsc{1}_\gamma - \Omega^\brsc{1}_b\right) - \frac{3}{8} R w_\gamma \Pi_B^\brsc{1}\; ,
\end{equation}
\end{subequations}
where $R = 4\rho_\gamma / 3 \rho_b$ and $\tau_c$ is the timescale for
Thomson scattering, the inverse of the opacity, $\tau_c^{-1} = a n_e
\sigma_T$. As with CDM there are no tensor perturbations to the baryon
distribution.

Describing the photon perturbations requires the full mechanics of the
Boltzmann distribution. Constructing the gauge invariant perturbation
equations is done in the same manner as for the neutrinos, with the
distinction that they are are massless bosons, and interact with the
baryons via Thomson scattering (see \cite{Durrer:1994,Hu:1997}). 
%Here
%we will neglect polarisation as we will not need the details of
%it. For calculating the effects on the CMB it is essential however, we
%recommend 
%\antcomment{so below equations are as if no polarization; maybe write in terms of a quadrupole source so at
%least consistent with numerical results?}
The full calculation requires a consistent treatment of polarization;
we do not repeat this here, see e.g.  Ref.~\cite{Hu:1997} for the
details. The photon hierarchy is concisely written as
\begin{multline}
\dot{\theta}_l^\brsc{m} = k \left[ \frac{\sqrt{l^2-m^2}}{2l-1}\theta_{l-1}^\brsc{m} - \frac{\sqrt{(l+1)^2-m^2}}{2l+3}\theta_{l+1}^\brsc{m}\right] \\- \theta_l^\brsc{m} / \tau_c + S_l^\brsc{m}
\end{multline}
the source terms $S_l^\brsc{m}$ describe the interactions with the
gravitational potentials and other matter species. The non-zero terms
are for the scalars
\begin{subequations}
\begin{align}
S_0^\brsc{0} & = \tau_c^{-1}\theta_0^\brsc{0} - \dot{\Phi} , & S_1^\brsc{0} & = \tau_c^{-1} V_b^\brsc{0} + k \Psi , \\
 S_2^\brsc{0} & = \tau_c^{-1} P^\brsc{0} , & & \notag
\end{align}
for the vectors
\begin{align}
S_1^\brsc{1} & = \tau_c^{-1} \Omega_b^\brsc{1} , & S_2^\brsc{1} & = -\frac{4}{\sqrt{3}} k \sigma^\brsc{1} + \tau_c^{-1} P^\brsc{1} ,
\end{align}
and for the tensors
\begin{equation}
S_2^\brsc{2} = \tau_c^{-1} P^\brsc{2} - \dot{H}_T
\end{equation}
\end{subequations}
where $P^\brsc{m}$ is the anisotropic Thomson source and contains the
coupling to the polarisation
\begin{equation}
P^\brsc{m} = \frac{1}{10}\left[\theta_2^\brsc{m} - \sqrt{6} E_2^\brsc{m} \right] \, .
\end{equation}
In terms of the photon multipole moments the usual matter sources are
\begin{align}
\Delta_\gamma & = \theta_0^\brsc{0} , & V_\gamma & = \frac{1}{4} \theta_1^\brsc{0} , & \Pi_\gamma^\brsc{0} & = \frac{3}{5} \theta_2^\brsc{0} , \notag \\
& &\Omega_\gamma^\brsc{1} & = \frac{1}{4} \theta_1^\brsc{0} , & \Pi_\gamma^\brsc{1} & = \frac{2\sqrt{3}}{5} \theta_2^\brsc{1} , \\
& & & &\Pi_\gamma^\brsc{2} & = \frac{2}{5} \theta_2^\brsc{2} \notag .
\end{align}

\subsection{Regular initial conditions}
We use the full system of perturbation equations to calculate initial
series solutions for 
modes well outside the horizon in the early radiation-dominated epoch, 
after neutrino decoupling but well before recombination.
%Such solutions are useful for
%analysing perturbations in the radiation dominated era, and are needed
These are needed
to provide the correct initial conditions for Boltzmann codes such as
\textsc{Camb} \cite{Lewis:2000} and \textsc{Cmbfast}
\cite{Zaldarriaga:1999}. We have calculated the complete set of all
known regular modes for the standard matter species (dark matter,
baryons, photons and neutrinos) for the scalar, vector and tensor type
perturbations. We have also included all the compensated magnetic
modes for three perturbations types. By using our expanded neutrino
equations (see Section~\ref{sec:massexpansion}) we can include
neutrinos of non-negligible mass, with solutions accurate to order
$m^2$. Thus our solutions include both massless neutrinos and a
number of degenerate massive species.

We make several standard approximations, firstly we assume we are in
the regime of tight coupling between photons and baryons where Thomson
scattering prevents slippage between the fluids, giving $V_b \approx
V_\gamma$ (see \cite{Ma:1995}). This gives two parameters which much
be small: there must be many scatterings per wavelength of the
perturbation $k \tau_c \ll 1$; and the scattering rate must be large
compared to the expansion rate $\tau_c / \tau \ll 1$. We take the
leading order corrections to this and truncate the tight coupling
hierarchy by assuming the photon anisotropic stress $\Pi_\gamma$ is
negligible
%\antcomment{is this true, I think scalar code uses the leading term for $\Pi_\gamma$?}
 (it is suppressed by a factor $k \tau_c$ relative to the
velocity). We also assume that the baryons are pressureless with $w_b
= c_{s,b}^2 = 0$, neglect any change in the background ionization fraction and degrees of freedom, 
and as before assume a flat universe. Standard dark energy does not affect the result until ${\cal O}(\tau^5)$.
%
%A deleted, not really an assumption (I think you comment on ignore octopole mode etc in the appendix)
%To close our neutrino hierarchy we assume that the $l =4$ and higher
%neutrino moments are zero. This should be valid in the early Universe
%as we expect successive moments to be smaller by factors of $k
%\tau$. 
%
%We also neglect the effects of both curvature, assuming a flat
%Universe as we have done all along, and that of any Dark Energy which
%(which in any standard form will only affect late times and orders higher
%than $\tau^5$).

The solutions are too lengthy to list in the main text and so we
include them in Appendix~\ref{app:initial}.

%%% Local Variables: 
%%% mode: latex
%%% TeX-master: "paper"
%%% End: 

\section{Primordial Magnetic Fields}

We will consider a stochastic background of magnetic fields
$B^i(x^j,\tau)$ generated by some mechanism in the very early
Universe. As for all the periods of interest the Universe contains a
highly ionized plasma, Maxwell's equations at first order show that
the field is frozen in, with an amplitude decaying with $1/a^2$. From
this we separate out the time evolution and write $B^i(x^j, \tau) =
B^i(x^j) / a(\tau)^2$. For a thorough discussion of the dynamics of
cosmological magnetic fields, see \cite{Barrow:2007}. The non-zero
components of the energy-momentum tensor are
\begin{subequations}
\begin{align}
T^0_0 & = - \frac{1}{8\pi a^4} B^2(\vx) \; , \notag\\
T^i_j & = \frac{1}{4\pi a^4} \left(\frac{1}{2} B^2(\vx) \delta^i_j - B^i(\vx) B_j(\vx)\right) \; .
\end{align}
\end{subequations}
As there is no magnetic field on the background, the perturbations of
the stochastic background are manifestly gauge invariant. We construct
two perturbations $\Delta_B$ and $\Pi_B$, defined by
\begin{subequations}
\begin{align}
T^0_0 & = - \rho_\gamma \Delta_B \; ,\notag\\
T^i_j & = p_\gamma \left(\Delta_B \delta^i_j + \Pi^{\phantom{B}i}_{B\phantom{i}j}\right) \; ,
\end{align}
\end{subequations}
where we include the factors of $\rho_\gamma$ and $p_\gamma$ to take
account of the $a^{-4}$ factors. As usual $\Pi^i_j$ can be decomposed
in the standard manner into scalar, vector and tensor contributions.

\subsection{Magnetic Modes}
\label{sec:magneticmodes}
Though the exact mechanism by which magnetic fields may be produced in
the primordial Universe is unclear, we are still able to address their
observational consequences. We imagine that the production of magnetic
fields occurs quickly at some time $\tau_B$, prior to the decoupling
of neutrinos from the photons at time $\tau_\nu$. We assume that this
decoupling is effectively instantaneous. Below we briefly review what
happens for the scalar case. This is discussed in detail in
\cite{Kojima:2009:2} using the synchronous gauge, where the
calculations are somewhat simpler. Our gauge-invariant notation has
the difficulty that some of the perturbations diverge on the
superhorizon scales we are interested in, and this needs to be
carefully addressed. The Mathematica notebook used for the
calculations of the gauge-invariant scalar, and tensor case can be
found at \url{http://camb.info/jrs/}, and describes these issues in
more detail.

Combining the four scalar Einstein equations \eqref{eq:einstein}
allows us to form the Bardeen equation for the potential $\Phi$ which
is sourced only by the total anisotropic stress $\Pi$ and the entropy
$\Gamma$
\begin{multline}
\label{eq:bardeen}
\ddot{\Phi} + 3\clh(1+c_s^2)\dot{\Phi} + [3(c_s^2 - w)\clh^2 +c_s^2 k^2]\Phi 
\\= 3 w \frac{\clh^2}{k^2} \Bigl[\frac{k^2}{2} \Gamma + \clh\dot{\Pi} -\frac{k^2}{3}\Pi \Bigr.\\\Bigl.+ 2\dot{\clh} \Pi + 3\clh^2\left(1 - c_s^2/w\right) \Pi \Bigr] \, .
\end{multline}
Prior to neutrino decoupling the Universe is dominated by the combined
radiative fluid with $c_s^2 = w = \frac{1}{3}$. In this limit the
Hubble parameter $\clh = \tau^{-1}$. The fluid is tightly bound to the
trace amount of Baryons and cannot develop any anisotropic stress, and
so the only anisotropic stress comes from the primordial magnetic
source, the constant $\Pi_B$. Until neutrino decoupling there is no
mechanism to compensate this, and it will act as a source for the
potentials. We will only discuss the anisotropic stress as the
magnetic density perturbation must be compensated at generation on
energy conservation grounds \cite{Stebbins:1990}. We reduce the
Bardeen equation to the radiation dominated limit
\begin{equation}
3 k^2\tau^2\left[\tau^2\ddot{\Phi} + 4 \tau \dot{\Phi}\right] + k^4 \tau^4 \Phi = -R_\gamma \Pi_B \left(6 + k^2 \tau^2\right)
\end{equation}
This can be solved exactly, and in the superhorizon limit of small
$k\tau$ it reduces to a solution of
\begin{equation}
\Phi(\tau) \approx \frac{R_\gamma \Pi_B}{k^2 \tau^2} -\frac{c_1}{k^3 \tau^3}  - \frac{c_1}{6 k \tau} + c_2 - \frac{2}{9} R_\gamma \Pi_B \log{(\tau)}
\end{equation}
which has a singularity for $k\tau = 0$. As we are concerned with
superhorizon modes, we check the physicality of this by examining the
co-moving curvature perturbation $\zeta = \Phi + 2 (\Psi + \dot{\Phi}
/ \clh) / 3(1 + w)$ finding that
\begin{equation}
\zeta(\tau) = \zeta(\tau_B) - \frac{1}{3}R_\gamma \Pi_B \left[\log{(\tau/\tau_B)} + \frac{\tau_B}{2\tau} -\frac{1}{2}\right]
\end{equation}
where we have absorbed the remaining constant terms by demanding
continuity of the $\zeta$ and the comoving density perturbation
(equivalent to continuity of $\Phi$). All the primordial contributions
to the curvature are contained within $\zeta(\tau_B)$.

At time $\tau_\nu$ the neutrinos decouple from the radiative fluid. By
considering their Boltzmann hierarchy we can examine what happens
next. Combining the $l = 1$ and $l = 2$ equations of
\eqref{eq:neutrino_eq} with the Bardeen equation \eqref{eq:bardeen} we
generate an equation for $\Pi_\nu$. As our gauge-invariant $\Delta$
and $V$ are divergent, we must carefully substitute them out. After
this we find a solution of the form
\begin{multline}
\Pi_\nu \approx -\frac{R_\gamma}{R_\nu} \Pi_B \biggl[1 - \sqrt{\frac{\tau_\nu}{\tau}} \Bigl[\cos{\bigl(\alpha \ln{(\tau/\tau_\nu)}\bigr)} \\+ d_1 \sin{\bigl(\alpha \ln{(\tau/\tau_\nu)}\bigr)}\Bigr]\biggr]
\end{multline}
where $\alpha$ is a positive constant depending on $R_\nu$. As $\tau
\rightarrow \infty$ we can see that the solution $\Pi_\nu \rightarrow
-\frac{R_\gamma}{R_\nu} \Pi_B$, compensating the magnetic anisotropic
stress. When the compensation is effective the source becomes zero and
the potentials stop growing. The further growth in the curvature can
be calculated giving the final curvature
\begin{equation}
\zeta \approx \zeta(\tau_B) -\frac{1}{3} R_\gamma \Pi_B \left[\log{(\tau_\nu/\tau_B)} + \left(\frac{5}{8 R_\nu} - 1\right)\right] \, ,
\end{equation}
where we have neglected terms in $\tau_B / \tau_\nu \ll 1$. 

Our initial conditions are given in the synchronous gauge and thus for
calculations we need the curvature perturbation in this gauge. It can
be calculated from $\zeta = \eta + \dot{\eta}/2\clh$ (in radiation
domination). On superhorizon scales, when the compensation is
complete, the derivative term will be zero, and $\eta(\tau) \approx
\zeta(\tau)$.

At some later time when the anisotropic stress is compensated their
are effectively two types of perturbation. The first is an
adiabatic-like mode with an amplitude $\zeta \sim - R_\gamma \Pi_B
\log{(\tau_\nu/\tau_B)} / 3$, the so-called passive mode, with all
species having zero initial anisotropic stress and unperturbed
densities. As we will see later, whilst the passive mode gives
adiabatic type perturbations, the statistics of $\Pi_B$ are
non-Gaussian unlike the standard adiabatic mode, and will have
significant higher order statistics \cite{Brown:2005}. The second type
is the well known compensated magnetic mode (see
\cite{Giovannini:2007,Finelli:2008,Mack:2002}), with no initial
curvature but containing the perturbed density and anisotropic
stresses (with the total density and anisotropic stress
unperturbed). We consider this in two parts: an anisotropic stress
sourced mode, with the compensating anisotropic stresses and
unperturbed densities, and a density sourced mode with unperturbed
anisotropic stresses but compensating densities. These have amplitudes
proportional to $\Pi_B$, and $\Delta_B$ respectively, and their
initial behaviour is presented in detail in
Appendix~\ref{app:initial}. Though we split them in two, these two
compensated modes are not independent; we address the statistics of
this in the next section.

The situation for the tensors is similar with, resulting in a passive
tensor mode of amplitude 
\begin{equation}
H^\brsc{2} \approx R_\gamma \Pi_B^\brsc{2} \left[\log{(\tau_\nu / \tau_B)} + \left(\frac{5}{8 R_\nu} - 1\right) \right]
\end{equation}
when the growth before and after decoupling is included. The
compensated mode is of amplitude $\Pi_B^\brsc{2}$. The vector mode has
no equivalent passive mode as perturbations purely to the vector
potential $\sigma^\brsc{1}$ decay away; it does have a compensated
mode, again of amplitude $\Pi_B^\brsc{1}$. For more details see
\cite{Lewis:2004}.

\subsection{Statistics}

The statistics of $B_i$ are assumed to be gaussian, and as we do not
include helical fields in our analysis \cite{Caprini:2009}, described
by a power spectrum $P_B(k)$ defined by
\begin{equation}
\left\langle B_i(\vk) B_j^*(\vk') \right\rangle = (2\pi)^3 \delta(\vk-\vk') \frac{P_{ij}(\khat)}{2} P_B(k)
\end{equation}
where $P_{ij} = \delta_{ij} - \khat_i \khat_j$ is a projection tensor
that comes from the zero divergence of $B$. Calculating the
energy-momentum perturbations requires us to consider them in harmonic
space, and this turns the real-space multiplications of $B$ into
$k$-space convolutions. This can then be used to calculate the power
spectra of the energy momentum perturbations in terms of convolutions
of the magnetic field power spectrum $P_B$. Various results have been
calculated for this, from approximations
\cite{Durrer:2000,Caprini:2002,Mack:2002} to exact results for
specific magnetic spectral indices \cite{Finelli:2008,Paoletti:2008}.
Since the energy momentum perturbations are quadratic in the magnetic
field, they cannot be Gaussian. Nonetheless the predicted power
spectrum is still interesting observationally, though more information
is available by also looking at higher-point
statistics~\cite{Brown:2005,Seshadri:2009,Caprini:2009b}.

Though there are two scalar magnetic sources, as they are both sourced
by the same underlying magnetic field, they are not independent, and
when considering their affect on the CMB we must carefully set up the
initial conditions for them with the correct amplitudes and
correlations between them, as well as the correct relative amplitude
of the vector and tensor contributions.

%To
%reason that this has little impact we can think of the perturbations
%as an approximate gaussian distributed field plus a homogenous
%contribution. This homogenous component can be taken as a negligible
%contribution to the background, and we can regard our analysis as
%pertaining to the remaining approximately gaussian component. For a
%thorough discussion of the non-gaussian statistics of the magnetic
%ield, including the three point function see \cite{Brown:2005}.

The scalar energy density perturbation is defined above. The scalar
anisotropic stress perturbation is $\Pi_B = -\frac{3}{2} T_{ij}(\khat)
\Pi^{ij}_B$ where we denote the traceless tensor $T_{ij}(\khat) =
(\khat_i \khat_j - \frac{1}{3}\delta_{ij})$. In terms of the magnetic
field these are written as
\begin{align}
\Delta_B &= \frac{1}{2} \delta_{ij} \Delta^{ij} \notag\\
\Pi_B & = \frac{9}{2} T_{ij}(\khat) \Delta^{ij}
\end{align}
where we have hidden the convolution of the magnetic field in a
definition of
\begin{equation}
\label{eq:delta_magdef}
\Delta^{ij} = \frac{1}{4\pi (2\pi)^3 \rho_\gamma a^4} \int d^3p\, d^3q\,  B^i(\vp) B^j(\vq) \delta(\vk - \vp - \vq) \;.
\end{equation}
There are three power spectra that we will need to compute, the power
spectra of both $\Delta_B$ and $\Pi_B$, and also, the oft-neglected
cross correlation of the two. In terms of two point statistics of
$\Delta^{ij}$
\begin{align}
\label{eq:magscal_ps}
\left\langle \Delta_B(\vk) \Delta_B^*(\vk') \right\rangle  &= \frac{1}{4} \delta_{ij} \delta_{lm} \left\langle \Delta^{ij}(\vk) \Delta^{lm *}(\vk') \right\rangle \, ,\notag\\
\left\langle \Delta_B(\vk) \Pi_B^*(\vk') \right\rangle  &= \frac{9}{4} \delta_{ij} T_{lm}(\khat') \left\langle \Delta^{ij}(\vk) \Delta^{lm *}(\vk') \right\rangle \,  ,\\
\left\langle \Pi_B(\vk) \Pi_B^*(\vk') \right\rangle  &= \frac{81}{4} T_{ij}(\khat) T_{lm}(\khat') \left\langle \Delta^{ij}(\vk) \Delta^{lm *}(\vk') \right\rangle \notag \, .
\end{align}
To calculate $\left\langle \Delta^{ij}(\vk) \Delta^{lm *}(\vk')
\right\rangle$ we substitute the definition \eqref{eq:delta_magdef},
and then using Wick's theorem to evaluate the 4-point correlator of
the gaussian $B$, we end up with a result in terms of a convolution of
$P_B$
\begin{multline}
\left\langle \Delta^{ij}(\vk) \Delta^{lm *}(\vk') \right\rangle =\\ \frac{\delta(\vk-\vk')}{16(2\pi)^2 \rho_\gamma^2 a^8} \int d^3p\, d^3q\, P_B(p) P_B(q) \delta(\vk-\vp-\vq) \\
\times \Bigl[ P^{il}(\phat) P^{jm}(\qhat) + P^{im}(\phat) P^{jl}(\qhat)\Bigr] \, .
\end{multline}
With this result we can calculate the power spectra of the scalar
perturbations by performing the relevant contractions of $T_{ij}$,
$P_{ij}$ and $\delta_{ij}$, which leave terms dependent on the angles
between $\khat$, $\qhat$ and $\phat$ (or $\widehat{(\vk-\vq)}$ as it will
become when we integrate out the Dirac-delta function). We will denote
$\gamma = \khat\cdot\qhat$, $\beta = \khat\cdot\phat$ and $\mu =
\phat\cdot\qhat$. The three correlations can be written in terms of
exact integrals, first
\begin{multline}
\label{eq:int_deltab}
\left\langle \Delta_B(\vk) \Delta_B^*(\vk') \right\rangle = \\ \frac{\delta(\vk-\vk')}{128\pi^2 \rho_\gamma^2 a^8} \int d^3q\, P_B(q) P_B(\lvert \vk -\vq\rvert) \left(1+\mu^2\right) \, ,
\end{multline}
second
\begin{multline}
\left\langle \Pi_B(\vk) \Pi_B^*(\vk') \right\rangle = \frac{9 \delta(\vk-\vk')}{32\pi^2 \rho_\gamma^2 a^8} \int d^3q\, P_B(q) P_B(\lvert \vk -\vq\rvert) \\ \times\left[1 - \frac{3}{4} (\gamma^2 + \beta^2) + \frac{9}{4} \gamma^2 \beta^2  - \frac{3}{2} \gamma\beta\mu + \frac{1}{4} \mu^2\right] \, ,
\end{multline}
and lastly the cross correlation
\begin{multline}
\left\langle \Delta_B(\vk) \Pi_B^*(\vk') \right\rangle = \frac{3 \delta(\vk-\vk')}{64\pi^2 \rho_\gamma^2 a^8} \int d^3q\, P_B(q) P_B(\lvert \vk -\vq\rvert) \\ \times\left[1 - \frac{3}{2} (\gamma^2 + \beta^2) + \frac{3}{2} \gamma\beta\mu - \frac{1}{2} \mu^2\right] \, .
\end{multline}
In the literature, the magnetic anisotropic stress $\Pi_B$ is often
replaced by the Lorentz force, given, in our notation, by $L_B =
\frac{2}{3}\left(w_\gamma \Pi_B - \Delta_B / 2\right)$. By combining
the correlations of $\left\langle \Delta_B(\vk) \Delta_B^*(\vk')
\right\rangle$ and $\left\langle \Delta_B(\vk) \Pi_B^*(\vk')
\right\rangle$ we can see that in general there is a non-zero
correlation between $L_B$ and $\Delta$, that has often been neglected
in the literature. It is given by
\begin{multline}
  \left\langle \Delta_B(\vk) L_B^*(\vk') \right\rangle =
  \frac{\delta(\vk-\vk')}{128\pi^2 \rho_\gamma^2 a^8} \int d^3q\,
  P_B(q) P_B(\lvert \vk -\vq\rvert) \\ \times\left[1 - 2 (\gamma^2 +
    \beta^2) + 2 \gamma\beta\mu - \mu^2\right] \, ,
\end{multline}
and should be included when calculating the effects that magnetic
fields have on the CMB.

We can calculate the relevant correlations for the vector and tensor
perturbations $\Pi_B^\brsc{1} = -6 k_{(i}e_{j)} ^\brsc{\pm 1}
\Delta^{ij}$ and $\Pi_B^\brsc{2} = -2 e_{ij}^\brsc{\pm 2} \Delta^{ij}$
in the same manner. The vector correlation is
\begin{multline}
\left\langle \Pi_B^\brsc{1}(\vk) \Pi_B^{\brsc{1} *}(\vk') \right\rangle = \frac{18 \delta(\vk-\vk')}{64\pi^2 \rho_\gamma^2 a^8} \int d^3q\, P_B(q) P_B(\lvert \vk -\vq\rvert) \\ \times\left[1 - 2 \gamma^2 \beta^2  + \gamma\beta\mu \right] \, ,
\end{multline}
and the tensor correlation is
\begin{multline}
\label{eq:int_pib_t}
\left\langle \Pi_B^\brsc{2}(\vk) \Pi_B^{\brsc{2} *}(\vk') \right\rangle = \frac{3 \delta(\vk-\vk')}{64\pi^2 \rho_\gamma^2 a^8} \int d^3q\, P_B(q) P_B(\lvert \vk -\vq\rvert) \\  \times(1 +  \gamma^2)(1 + \beta^2)\, .
\end{multline}
Our results are in agreement with those in the literature
\cite{Brown:2005,Paoletti:2008,Caprini:2002}.

The exact form of the magnetic power spectrum $P_B(k)$ is highly
dependent on the production mechanism. We follow the rest of the
literature in choosing to use a power law description
\begin{equation}
P_B(k) = A k^{n_B}
\end{equation}
for $k < k_D$, and zero otherwise. The cutoff wavenumber $k_D$ comes
from the fact that radiation viscosity leads to damping of small scale
magnetic fields. This is the order of the Silk-damping scale times the
dimensionless Alfv\'{e}n-velocity
\cite{Jedamzik:1998,Subramanian:1998}, which is time dependent, though
we are mainly interested in perturbations sourced around and before
recombination.  The amplitude $A$ is defined in terms of the expected
field amplitude $B_\lambda$ smoothed on a scale $\lambda$ (we use the
conventional $\lambda = 1 \mathrm{Mpc}$). This gives
\begin{equation}
  A = \frac{(2\pi)^{n_B + 5} B_\lambda^2}{\Gamma\left(\frac{n_B+3}{2}\right) k_\lambda^{n_B + 3}} \, .
\end{equation}
For illustration we shall focus on nearly scale-invariant magnetic
field spectra, since these are the only ones likely to give signals in
the CMB on acoustic-oscillation scales
\cite{Caprini:2002,Durrer:2003}. It should be noted that it is
difficult for causal mechanisms to give such spectra, and so to
produce large scales modes we are likely to need some inflationary
mechanism.

For scale-invariant spectra the contributions of interest are then
from scales much larger than the damping scale $k_D$, for a spectral
index $n_B < -3/2$, the cutoff becomes largely irrelevant. The effect
on the $C_l$'s from modifying the power spectrum at these scales is
small. For the compensated modes it is around $1$ percent at $l \sim
2000$, and less than $3$ percent at $l \sim 5000$. The effect on the
passive modes will be negligible as the magnetic damping scale is tiny
at neutrino decoupling.

Ignoring the cutoff in the definitions of $P_B$ allows us factor out
the $k$-dependence of the above integrals and make them dimensionless,
depending only on the spectral index. For instance the integral in
\eqref{eq:int_deltab} can be rewritten as
\begin{multline}
\int d^3q\, P_B(q) P_B(\lvert \vk -\vq\rvert) \left(1+\mu^2\right) = 2\pi k^{2 n_B + 3} \\ \times\int^\infty_0 \!\!\! \int_{-1}^1\!\! du\, d\gamma \: u^{n_B} \left(1 - 2 u \gamma + u^2\right)^{n_B /2} (1 +\mu^2)
\end{multline}
where we have substituted $u = q/k$. The angular functions $\mu$ and
$\beta$ can be written in terms of $\gamma$ and $u$ as
\begin{align}
\mu & = \qhat\cdot\widehat{(k-q)} = \frac{\gamma - 1}{(1 - 2 u \gamma + u^2)^{1/2}} \, ,
\\
\beta & = \khat\cdot\widehat{(k-q)} = \frac{1 - \gamma u}{(1 - 2 u \gamma + u^2)^{1/2}} \, .
\end{align}
The same can be done for all the correlations above
\eqref{eq:int_deltab}--\eqref{eq:int_pib_t}. Whilst the integrands
have singularities at $u = 0$ (corresponding to $\vq = 0$), and $u =
1$, $\gamma = 1$ (corresponding to $\vk - \vq = 0$), the integrals are
convergent provided that $n_B > -3$. We use a series expansion to
integrate small regions around each of the poles, and numerically
integrate the remainder. We use a nearly scale invariant power spectrum
with $n_B = -2.9$ giving power spectra
\begin{align}
P_{\Delta_B}(k) &= \frac{(53.29)}{4}\: \left[\frac{(2\pi)^{n_B +2}}{2\Gamma\left(\frac{n_B +3}{2}\right)}\frac{B_\lambda^2}{\rho_{\gamma 0}}\right]^2\left(\frac{k}{k_\lambda}\right)^{2 n_B + 6} \notag\\
P_{\Delta\Pi_B}(k) &= -\frac{3(25.93)}{2}\: \left[\frac{(2\pi)^{n_B +2}}{2\Gamma\left(\frac{n_B +3}{2}\right)}\frac{B_\lambda^2}{\rho_{\gamma 0}}\right]^2\left(\frac{k}{k_\lambda}\right)^{2 n_B + 6} \notag\\
P_{\Pi_B}(k) &= 9(14.55) \: \left[\frac{(2\pi)^{n_B +2}}{2\Gamma\left(\frac{n_B +3}{2}\right)}\frac{B_\lambda^2}{\rho_{\gamma 0}}\right]^2\left(\frac{k}{k_\lambda}\right)^{2 n_B + 6} \notag\\
P_{\Pi_B}^\brsc{1}(k) &= 9(26.30) \: \left[\frac{(2\pi)^{n_B +2}}{2\Gamma\left(\frac{n_B +3}{2}\right)}\frac{B_\lambda^2}{\rho_{\gamma 0}}\right]^2\left(\frac{k}{k_\lambda}\right)^{2 n_B + 6} \notag\\
P_{\Pi_B}^\brsc{2}(k) &= \frac{3(105.55)}{2} \: \left[\frac{(2\pi)^{n_B +2}}{2\Gamma\left(\frac{n_B +3}{2}\right)}\frac{B_\lambda^2}{\rho_{\gamma 0}}\right]^2\left(\frac{k}{k_\lambda}\right)^{2 n_B + 6} \notag\\
\end{align}
%\antcomment{fix power spectrum def}
where our power spectra are defined in a dimensionless manner
$\left\langle \Delta_B(\vk) \Delta_B^*(\vk') \right\rangle = 2\pi^2
\left(2\pi\right)^3 \delta(\vk - \vk') k^{-3} P_{\Delta_B}(k)$. The
numerically calculated value is wrapped in parentheses. Note that
these power spectra only include one of the two separate modes for the
vector and tensor type perturbations. The shape of our power spectra
are identical to the commonly used approximations of \cite{Mack:2002},
but our integration predicts significantly different amplitudes. Using
the same approximation scheme as \cite{Mack:2002} we would predict
that the angular integrals are equal to $2 n / ( n + 3 ) ( 2 n + 3
)$. For our spectral index this is approximately $\sim 20.7$.
Comparing this to the numerical results above (shown in parentheses),
we see that the difference is up to around five times for the tensor
power spectrum.

\begin{figure}[tb]
\hspace{-0.05\linewidth}\includegraphics[width=0.95\linewidth]{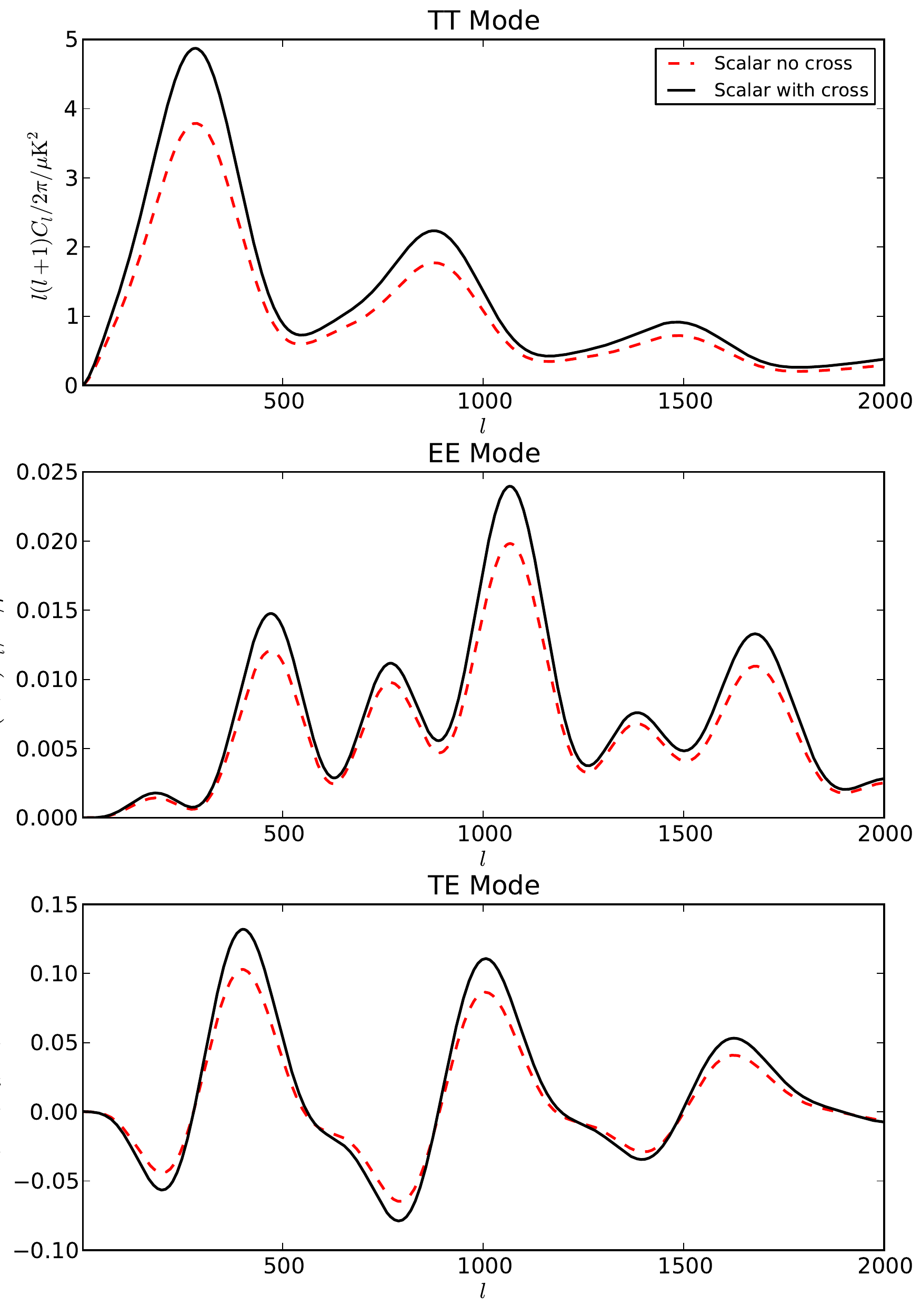}
\caption{The scalar power spectra with and without the
  cross-correlation between $\Delta_B$ and $\Pi_B$. Inclusion of it in
  calculations gives a consistent increase in power of around 15--25
  percent at all scales.}
\label{fig:cmb_cross}
\end{figure}

In Figure~\ref{fig:cmb_cross} we show the effect that the
cross-correlation between $\Delta_B$ and $\Pi_B$ has on the
CMB. Despite it being an anti-correlation we can see that it boosts
power on all scales, as many of the perturbations are effectively
sourced by the Lorentz force $L_B = \frac{2}{3}\left(w_\gamma \Pi_B -
  \Delta_B / 2\right)$.

% Note that to get to Antony's result we need to
%a) change out of gaussian results (multiply by (4\pi)^2)
%b) apply correction for the vector (divide by 2) and tensor (multiply by 3) harmonics

% \begin{align}
% \left\langle \Delta_B(\vk) \Delta_B(\vk') \right\rangle &= 4.22\times 10^{-2}\:  \frac{\delta(\vk+\vk')}{\rho_{\gamma 0}^2} A^2 k^{2 n_B + 3} \notag\\
% \left\langle \Delta_B(\vk) \Pi_B(\vk') \right\rangle &= -0.123\:  \frac{\delta(\vk+\vk')}{\rho_{\gamma 0}^2} A^2 k^{2 n_B + 3} \notag\\
% \left\langle \Pi_B(\vk) \Pi_B(\vk') \right\rangle &= 0.415\: \frac{\delta(\vk+\vk')}{\rho_{\gamma 0}^2} A^2 k^{2 n_B + 3}\\
% \left\langle \Pi_B^\brsc{1}(\vk) \Pi_B^\brsc{1}(\vk') \right\rangle &= 0.749\:  \frac{\delta(\vk+\vk')}{\rho_{\gamma 0}^2} A^2 k^{2 n_B + 3} \notag\\
% \left\langle \Pi_B^\brsc{2}(\vk) \Pi_B^\brsc{2}(\vk') \right\rangle &= 0.501\:  \frac{\delta(\vk+\vk')}{\rho_{\gamma 0}^2} A^2 k^{2 n_B + 3} \notag
% \end{align}

\subsection{Numerical Calculation}

\begin{figure*}[tb]
\hspace{-0.05\textwidth}\includegraphics[width=0.95\textwidth]{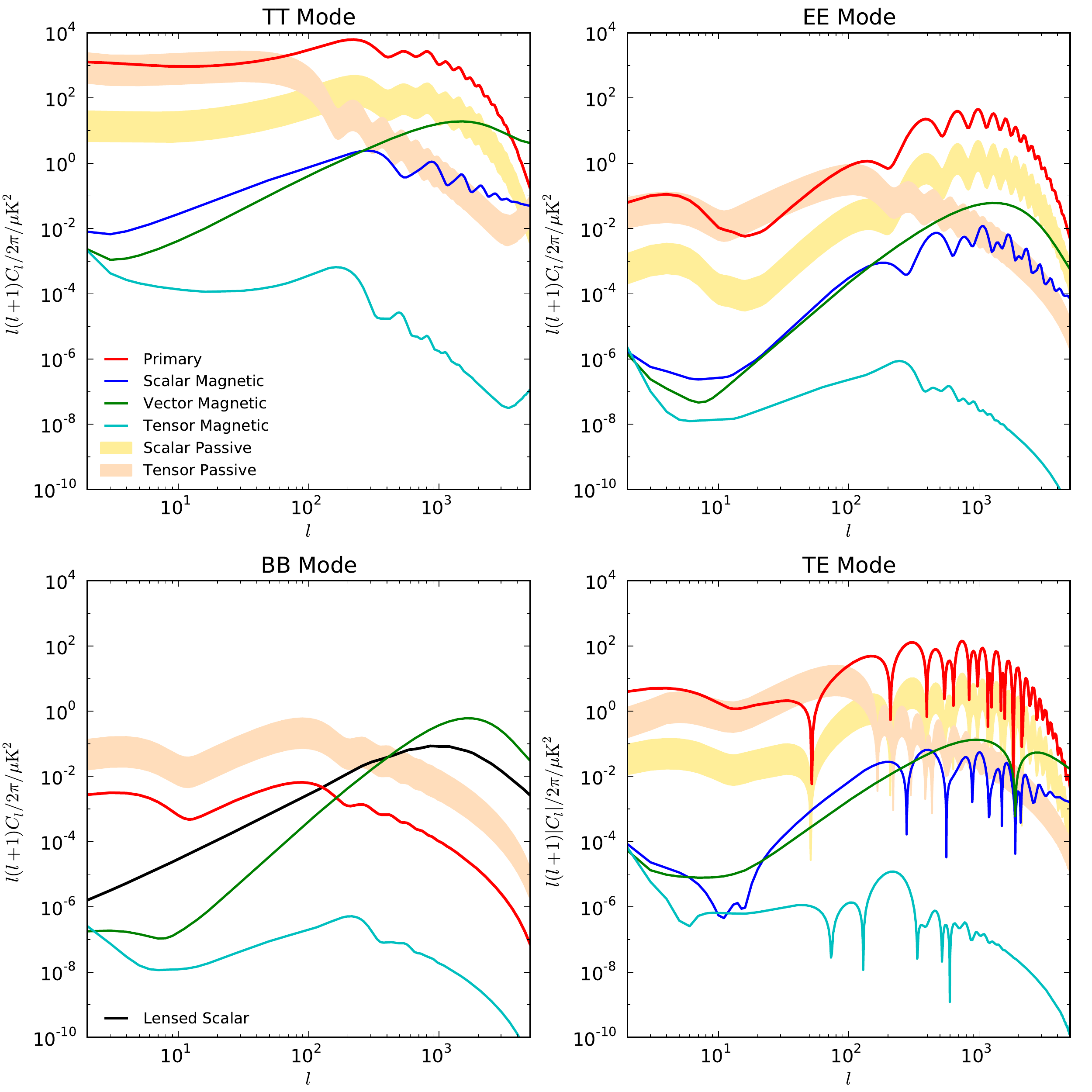}
\caption{The four CMB power spectra plotted for a realistic neutrino
  mass $\sum m_\nu = 0.47 \:\mathrm{eV}$, with a magnetic field
  $B_\lambda = 4.7 \:\mathrm{nG}$. We include the scalar primary
  contribution for the TT,EE and TE power spectra, and the tensor
  primary (with a tensor to scalar ratio of 0.1) and for the BB power
  spectrum. The shaded regions represent the regions we would expect
  the passive modes to lie within for production between the reheating
  and the electroweak transition.}
\label{fig:cmb_realistic}
\end{figure*}

In Figure~\ref{fig:cmb_realistic} we plot the four CMB power spectra
for primordial magnetic fields. We use the constraint of
\cite{Yamazaki:2006}, of $B_\lambda = 4.7 \:\mathrm{nG}$ at a scale of
$\lambda = 1 \: \mathrm{Mpc}$, with a realistic neutrino mass $\sum
m_\nu = 0.47 \:\mathrm{eV}$ taken from the recent constraints of
\cite{Reid:2009}. We include both the compensated modes for all three
perturbation types as well as the passive modes. Note that within this
paper we assume that the magnetic perturbations are uncorrelated with
the primary sources of anisotropy in the CMB.

There is currently no leading theory of the formation of primordial
magnetic fields, though there is much work suggesting their production
could be around the electroweak phase transition \cite{Baym:1996} at
$T \sim 1 \:\mathrm{TeV}$, or from just after the Quark-Hadron phase
transition \cite{deSouza:2008,deSouza:2009} at $T \sim 150
\:\mathrm{MeV}$. However, to produce a scale-invariant spectrum we are
likely to need some kind of acausal inflationary method
\cite{Bertolami:1998}, though many often struggle to produce large
enough magnetic fields. For an unknown inflationary mechanism the
exact time and details of magnetic field production are unclear,
however, for illustration we believe that the electroweak transition
provides a useful bound on the latest production time, and reheating
(at temperature $T < 10^{14} \mathrm{GeV}$, about the GUT scale) a
bound on the earliest. This gives $\tau_\nu / \tau_B \sim
10^6$--$10^{12}$. Any magnetic perturbations directly generated during
inflation will source passive modes which are essentially just a
component of the primordial spectra.
%\antcomment{how do you get nearly scale invariant from electroweak?}

Our CMB power spectrum results are in broad agreement with previous
work \cite{Paoletti:2008,Lewis:2004} in the cases where the results
have been calculated. 

The most significant magnetic contributions to the CMB come from the
tensor passive modes in all four power spectra. This is at the level
of 10 percent for the temperature anisotropies and around an order of
magnitude greater than the primordial gravitational wave contribution
to the B-mode polarisation.  The compensated vector mode is important
on very small scales in the temperature power spectrum
\cite{Subramanian:2002}, though here we also have to cope with
significant secondary contributions from SZ\footnote{Recent work has
  suggested a strong constraint from the magnetic mode contribution to
  the SZ effect \cite{Tashiro:2009}} and CMB lensing. The vector mode
also leaves a clear signature in the B-mode polarization spectrum on
small scales, with a comparable amplitude but different shape to the
secondary signal expected from CMB lensing.

Its large amplitude at low multipoles mean that the passive mode may
provide stronger constraints on any primordial magnetic field than the
compensated mode, though the relative amplitude between the two is
uncertain due to the unknown epoch of magnetic field production (but
the dependence is only logarithmic). Using current WMAP temperature
data, the passive modes should constrain the magnetic field to lower
than the current linear-theory CMB-only limit $B_\lambda < 4.7
\:\mathrm{nG}$ of \cite{Yamazaki:2006}. Planck B-mode data will only
enhance this. The effectiveness of these CMB constraints will be
limited by: secondary effects at small scales obscuring the
compensated vector mode; and confusing primordial tensor modes with
the passive modes on large scales (with a large cosmic variance). We
should also note that as the amplitude of the power spectra scales
like $B_\lambda^4$, improving the upper limits on the magnetic modes,
translates into much weaker improvements in the magnetic field
strength constraints.

The results of \cite{Kojima:2008} suggested that the presence of
massive neutrinos led to a significant enhancement in power in the
compensated modes at the largest scales. Whilst we see an see an
increase in power on large scales due to massive neutrinos, the effect
we calculate is much less significant (by about five orders of
magnitude). We believe this effect is due to a numerical issue with
tight-coupling which we discuss in detail in Section~\ref{sec:tight}.

%\input{magneticmodes}
%\section{Results}

\section{Numerical Issues}

\subsection{Tight Coupling}

\label{sec:tight}
To derive a tight coupling approximation for tensors, we take the
evolution equations for the CMB temperature and E-mode polarization
quadrupole
\begin{equation}
\dot{\theta}_2^\brsc{2} = - k \frac{\sqrt{5}}{7} \theta_3^\brsc{2} - \frac{9}{10} \tau_c^{-1} \theta_2^\brsc{2} -\frac{\sqrt{6}}{10} E_2^\brsc{2} - \dot{H}^\brsc{2} \; ,
\end{equation}
and
\begin{equation}
\dot{E}_2^\brsc{2} = -k\left[\frac{2}{3} B_2^\brsc{2} + \frac{5}{2} E_3^\brsc{2}\right] - \frac{2}{5\tau_c} \left[ E_2^\brsc{2} + \frac{\sqrt{6}}{4} \theta_2^\brsc{2}\right] \; .
\end{equation}
To obtain an equation for $\theta_2^\brsc{2}$, we rearrange these two
equations, and substitute for $E^\brsc{2}_2$ to give
\begin{multline}
\label{eq:tc1}
\theta_2^\brsc{2} = k \tau_c \left[\frac{3\sqrt{6}}{10} E_3^\brsc{2} + \frac{2\sqrt{6}}{25} B_2^\brsc{2} - \frac{4}{3} \theta_3^\brsc{2}\right] \\+ \tau_c \frac{d}{d\tau} \left[\frac{6\sqrt{6}}{50} E_2^\brsc{2} - \frac{4}{3} \theta_2^\brsc{2} - \frac{4}{3} H^\brsc{2}\right] \; .
\end{multline}
Looking at this, we see that for small $k \tau_c$ and small $\tau_c /
\tau$, the temperature quadrupole is also small, though we don't show
it this is also true for the E-mode quadrupole. Physically this can be
interpreted as the photons are tightly coupled to the baryons if there
are many scatterings within a wavelength of the perturbation, and
there are many scatterings across the horizon size. Rearranging the
equation for higher temperature moments
\begin{multline}
\theta_l^\brsc{2} = k\tau_c \biggl[ \frac{\sqrt{(l-1)^2-4}}{2 l -1} \theta_{l-1}^\brsc{2} \\- \frac{\sqrt{(l+1)^2-4}}{2 l + 3} \theta_{l+1}^\brsc{2}\biggr] - \tau_c \frac{d}{d\tau} \theta_l^\brsc{2}
\end{multline}
we see that higher moments are suppressed by factors of $k \tau_c$,
that is $\theta_l^\brsc{2} \propto k\tau_c \theta_{l-1}^\brsc{2}$. If
we want to only retain terms up to first order in $\tau_c$, this
allows us to ignore higher moments in \eqref{eq:tc1}. Noting that
$B_2^\brsc{2} \propto k\tau_c E_2^\brsc{2}$ by the same argument, we
can drop all terms $\propto k\tau_c$ leaving
\begin{equation}
\label{eq:tc2}
\theta_2^\brsc{2} = \tau_c \frac{d}{d\tau} \left[\frac{6\sqrt{6}}{50} E_2^\brsc{2} - \frac{4}{3} \theta_2^\brsc{2} - \frac{4}{3} H^\brsc{2}\right] \; .
\end{equation}
If both $k\tau_c$ and $\tau_c / \tau$ are small, then the
$E_2^\brsc{2}$ and $\theta_2^\brsc{2}$ terms in the right hard bracket
are small corrections to the value of overall $\theta_2^\brsc{2}$ and
we can neglect them leaving
\begin{equation}
\theta_2^\brsc{2} = -\frac{4}{3} \tau_c \dot{H}^\brsc{2} \; ,
\end{equation}
the standard tight-coupling approximation for the tensors.

The problem within CAMB (Feb 2009 version) is that for tensor modes
with small $k\tau_c$ it uses the tight coupling approximation no
matter what the value of $\tau_c / \tau$. This clearly invalidates the
tight coupling approximation, but it does not manifest itself for
standard models as most of the quantities we are interested in are
proportional to $k$ anyway, and thus the overall error is small
(generally much smaller than 1 percent on the largest scale $C_l$'s).

As can be seen from the initial conditions in
Appendix~\ref{app:initial}, the growth of modes like the tensor
compensated magnetic mode is modified and they grow proportional to
$k_\text{eff}^2 \tau^2$ with an effective wavenumber $k_{\text{eff}}^2
= k^2 + \alpha \bar{m}^2$ and thus on very large scales
$k_\text{eff}^2 \propto \bar{m}^2$. This degenerate evolution ensures
that the growth of large scale perturbations remains large, and thus
there is a large error from the tight coupling approximation.

\begin{figure}[tb]
\hspace{-0.05\linewidth}\includegraphics[width=1.05\linewidth]{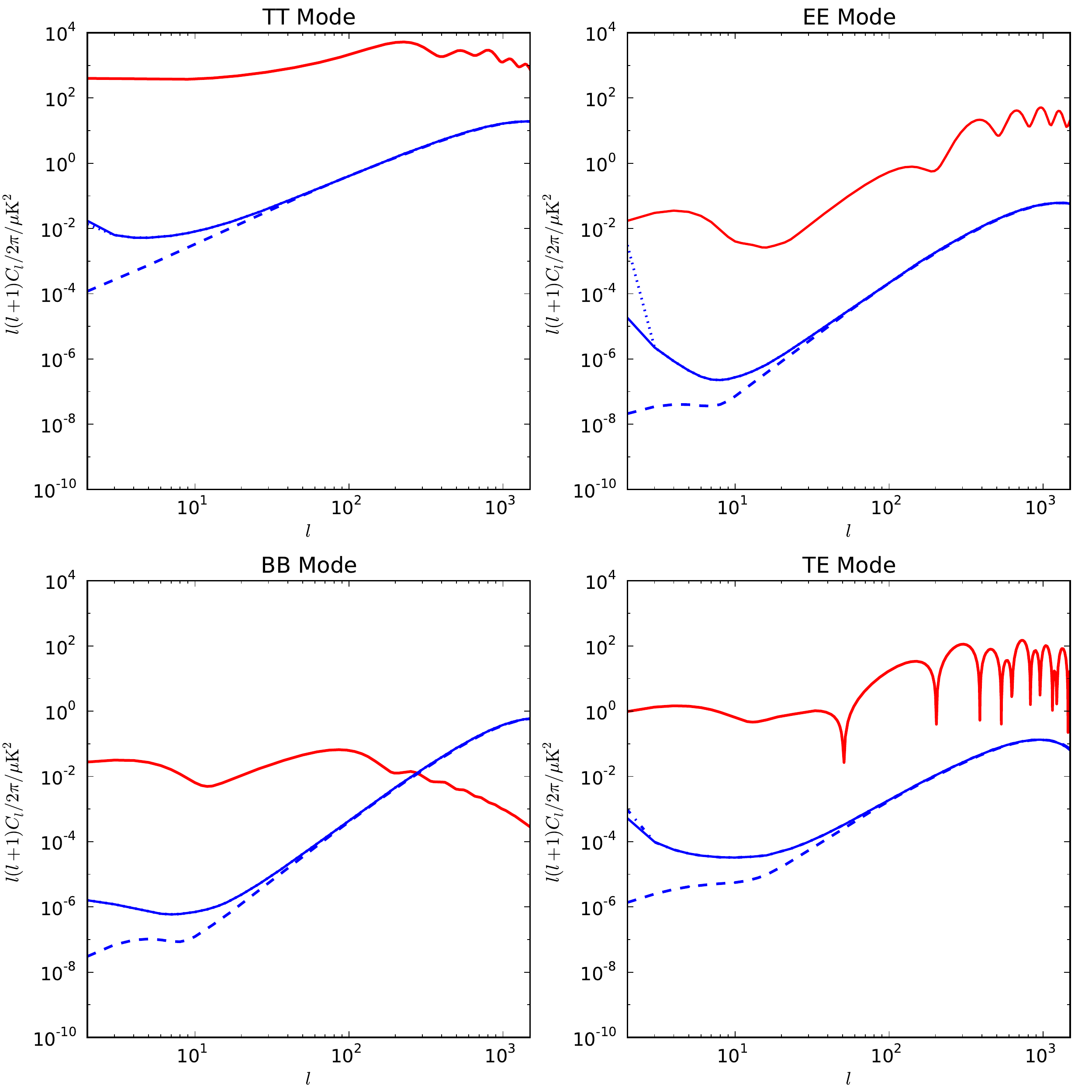}
\caption{The compensated vector contributions to angular power spectra
  of the temperature and polarization of the CMB. For each spectrum we
  plot three different cases, for purely massless neutrinos (dashed),
  and for massive neutrinos ($\sum m_\nu = 1.8 \mathrm{eV}$)
  calculated using the CAMB defaults (dotted), or our modified version
  (solid). In all cases we use a magnetic field of $B_\lambda = 4.7
  \mathrm{nG}$. We also include the primary contribution to the
  spectrum in each case (thick solid), scalar perturbations for the
  TT, EE, TE plots, and the gravitational wave contribution to
  BB. Whilst both massive neutrino cases contain significant large
  scale power compared to the massless neutrinos, our modifications
  avoid the artificial increase at very low $l$ given by the CAMB
  default.}
\label{fig:cmb_vector}
\end{figure}

\begin{figure}[bt]
\hspace{-0.05\linewidth}\includegraphics[width=1.05\linewidth]{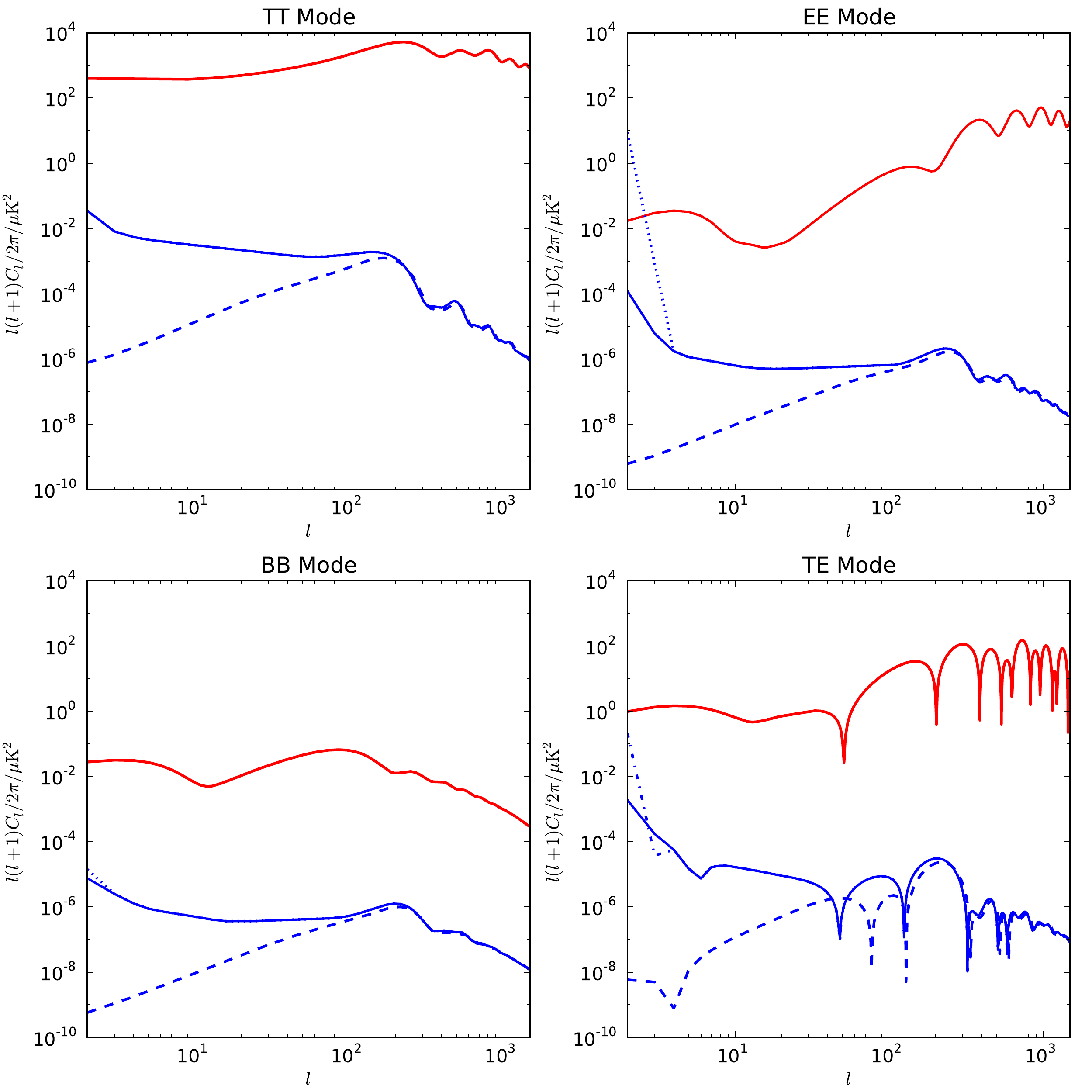}
\caption{The compensated tensor mode to the four CMB angular power
  spectra of temperature and polarization. This is the tensor
  equivalent of Fig.~\ref{fig:cmb_vector}. The solid line is our
  modified version, dotted the CAMB default and dashed the massless
  case. The CAMB default exhibits the same small $l$ excess as in the
  vector case, and as before our modified version avoids this.}
\label{fig:cmb_tensor}
\end{figure}

In \figref{fig:cmb_vector} and \figref{fig:cmb_tensor} we show the
vector and tensor contributions to the four CMB power spectra before
and after correcting the tight coupling behaviour. As we can see this
has significant effects, most notably on the tensor contribution to
the EE mode power spectrum, compared to the default behaviour of
CAMB. This explains the tremendous increase in large scale $E$-mode
power seen in the results of \cite{Kojima:2008} where we have used the
same total neutrino mass $\sum m_\nu = 1.8 \textrm{eV}$, magnetic
field strength $B_\lambda = 4.7\,\textrm{nG}$ and magnetic spectral
index $n_B = -2.9$. Our calculation shows that, even at the lowest
multipoles, the tensor compensated magnetic mode is significantly
lower amplitude than the primary scalar adiabatic spectrum, and remain
subdominant to the compensated vector mode.

\subsection{Early-Time Numerical Instabilities}

\begin{figure}[tb]
\hspace*{-0.09\linewidth}\includegraphics[width=1.1\linewidth]{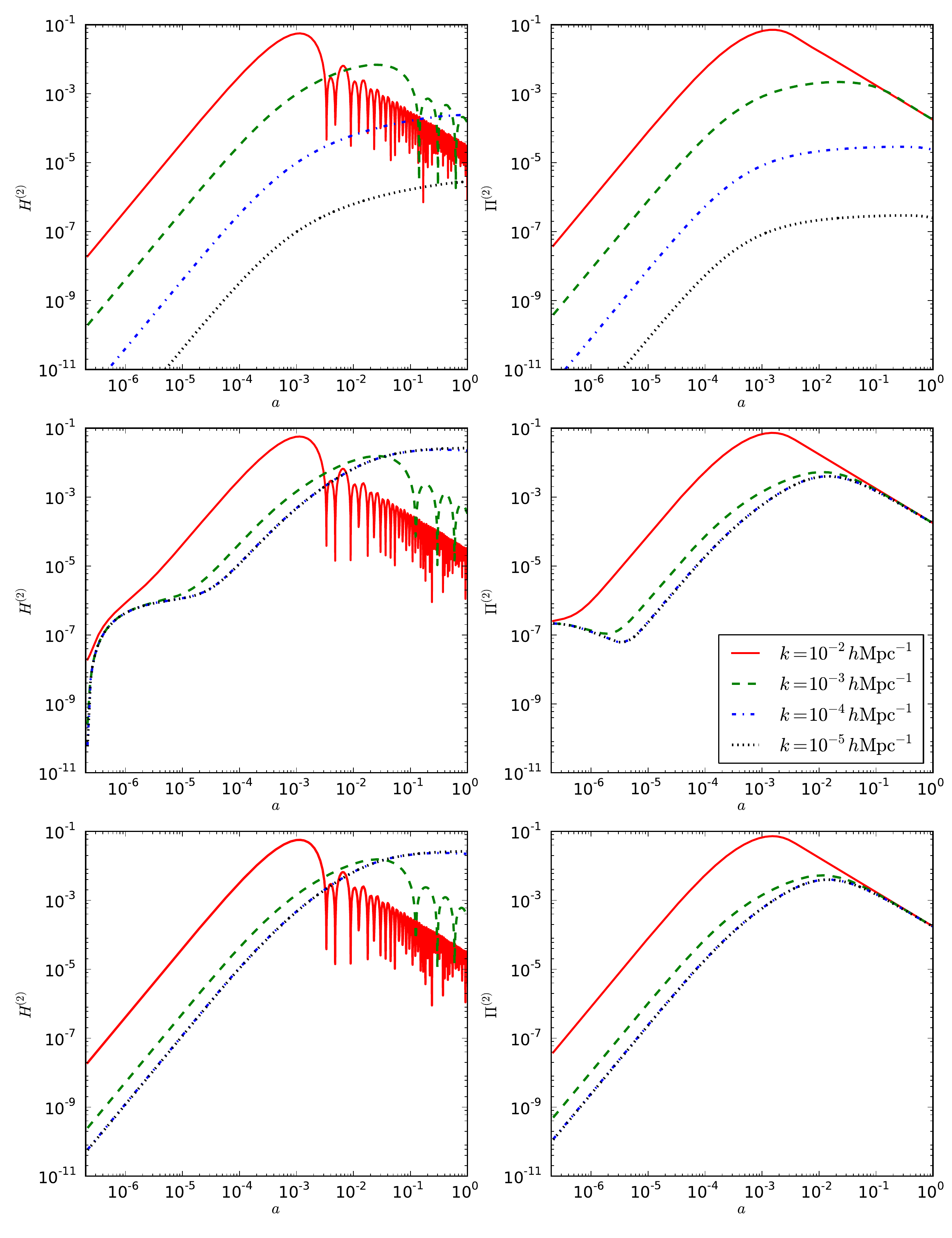}
\caption{The evolution of the tensor metric perturbation $H^\brsc{2}$
  (left panels), and the total anisotropic stress $\Pi^\brsc{2}$
  plotted against the scale factor $a$, at various wavenumbers. In the
  top panel we show the evolution with massless neutrinos. The middle
  panels illustrate the behaviour when we instead use three massive
  neutrinos $\sum m_\nu = 0.18 \mathrm{eV}$, with the default behaviour
  of CAMB. The problems stemming from the integration accuracy are
  readily apparent at early times. The bottom panels show the correct
  evolution of the massive neutrinos with our modifications. The
  degenerate evolution at small $k$ is apparent.}
\label{fig:tensor_evo}

\end{figure}

In Fig.~\ref{fig:tensor_evo} we show the evolution of the tensor
perturbations of several different large scales for the compensated
magnetic mode. The top set of panels show the behaviour in the
presence of massless neutrinos, and we see the slowly growing
scale-dependent evolution of both the gravitational waves and the
total anisotropic stress. The middle panels show the output of CAMB
when evolving massive neutrinos, illustrating a fundamental problem
when numerically evolving these perturbations. To evolve neutrino
quantities such as the anisotropic stress, we need to evolve the
distribution function perturbation $\Psi_\nu(k_i, q, n^j)$ at a fixed
set of points $q$, then we numerically integrate over the points to
calculate the desired quantity. For the standard modes this approach
is fine, however, in the case of the compensated magnetic mode, the
initial cancellation is at the order of $10^{-10}$, and require
numerical accuracy at this level to calculate the anisotropic stress
correctly. As well as simple numerical accuracy, we must integrate
well into the tail of the distribution to include all contributions up
to $10^{-10}$, this requires an increase in the range of $q$ values
integrated from $q_\text{max} \sim 15 k_B T$ up to around
$q_\text{max} \sim 40 k_B T$.

One way to obtain the numerical accuracy would be to simply increase
the number of points over which we integrate. However, combined with
the required increase in range, this requires a significant increase
in the number of integration points. We use an alternative approach,
using our mass expansion of the Boltzmann hierarchy. At early times we
directly evolve the integrated moments $F_l$ and use this to calculate
the anisotropic stress. As the neutrinos start to become
non-relativistic, our mass expansion becomes inaccurate and so before
this we switch to using the full distribution function. By this time
the level of cancellation is within the numerical accuracy of the
integration and the total anisotropic stress is accurate. The results
of this are shown in Fig.~\ref{fig:tensor_evo}.

Such an approach is essential to accurately model the behaviour of
massive neutrinos in the early Universe, however for calculating CMB
power spectra the correctons are sub-percent level and simply
increasing the range and number of integration points is sufficient.

\section{Conclusion}

In this paper we have developed an integrated Boltzmann hierarchy for
analysing massive neutrinos in the early Universe which is accurate to
second order in the mass. We have calculated the leading
order mass corrections to the initial series solutions for the regular
perturbation modes, and also demonstrated its use for accurately
evolving massive neutrinos in the early Universe.

We have made a detailed analysis of the effects of the primordial
magnetic fields on the CMB. In our examination of the statistics of
the magnetic field perturbations we have included an often neglected
cross-correlation term between the two scalar-perturbations. This
serves to increase the power in the CMB from the compensated mode by
around 25 percent at all scales. We also demonstrate that one of the
standard approximations to the statistics can give an amplitude around
a factor of five smaller than a more accurate result, reinforcing the
need to move to more advanced results.

We accurately calculate the contributions of the various magnetic
modes (both passive and compensated to the CMB). By correcting some
numerical issues we come to different conclusions to
\cite{Kojima:2008}. Whilst we agree that there is an enhancement to
the large scale power spectra (especially E-mode polarization) caused
by massive neutrinos, we find a much smaller amplitude; too small to
enhance prospects of detecting primordial magnetic fields. Our work
suggests that the passive modes are likely to dominate the compensated
modes with a power spectrum amplitude several orders of magnitude
greater at large scales. With the magnetic field we have used they are
around 10 percent of the primary spectra, and this suggests that they
will provide the biggest constraint on any primordial magnetic field
in the near future, adding in a small gravitational wave like
component with a blue spectral index. However unlike the compensated
mode, such modes are dependent on the details of the magnetic field
production, though quite weakly, and cannot provide model independent
constraints on the magnetic field in the same manner as the
compensated modes.

The Mathematica notebooks to calculate the magnetic mode amplitudes,
and the initial conditions can be found at
\url{http://camb.info/jrs/}. Our modifications to CAMB to calculate the
compensated magnetic modes are at the same location.
\nopagebreak[4]\subsection*{Acknowledgements} We would like to thank
Anthony Challinor for useful discussion. JRS is supported by an STFC
Studentship, and AL by an STFC Advanced Fellowship.

\appendix
\newcommand{\ud}{{\rm d}}
\newcommand{\begm}{\begin{pmatrix}}
\newcommand{\enm}{\end{pmatrix}}

\newcommand{\md}{{\widetilde{m}}}

\section{Higher-order mass expansion}
\label{app:higherorder}
\noindent We define a scaled mass
\begin{equation}
\md = a \frac{m}{k_B T_0}
\end{equation}
so the ratio of massive and massless neutrino densities is given by
\begin{equation}
\frac{\rho_\nu}{\rho_{\nu 0}} = \frac{120}{7\pi^4} I(\md)
\end{equation}
where
\begin{equation}
I(\md) \equiv \int_0^\infty \ud q q^2 \frac{\sqrt{q^2+\md^2}}{e^q + 1}.
\end{equation}
Performing an expansion of $I(\md)$ in the mass by performing a series
expansion of the square root inside the integral is not valid since
$\md$ is not much smaller than $q$ over the full range of the
integral. Instead we split up the integral at a point $\alpha$ (where
$\md \ll \alpha \ll 1$) so that
\begin{equation}
I(\md) = \frac{7}{120}\pi^4 + \frac{\pi^2}{24} \md^2 + I_1(\md) + I_2(\md) + I_3(\md)
\end{equation}
where
\begin{eqnarray}
I_1(\md) &\equiv& \int_0^\alpha \ud q q^2 \frac{1}{2}\left[\sqrt{q^2+\md^2}- q^3 - \md^2 q/2\right]
\\&=&
 \frac{\alpha}{16}\sqrt{\md^2+\alpha^2}(\md^2 + 2\alpha^2) - \frac{\md^4}{16} \sinh^{-1}\left(\frac{\alpha}{\md}\right) \nonumber
 \\ &&- \frac{\alpha^2}{8}(\alpha^2+\md^2).\nonumber
 \\ &=& \frac{\md^4\ln \md}{16}  + \left( 1 - 4 \ln(2\alpha)\right) \frac{\md^4}{64} - \frac{\md^6}{64\alpha^2} + {\cal O}(\md^8) \nonumber
\end{eqnarray}
\begin{align}
&I_2(\md) \equiv \\
&\int_0^\alpha \ud q \left(q^2 \sqrt{q^2+\md^2} -q^3 - \md^2 q/2\right)\left( \frac{1}{e^q+1} - \frac{1}{2}\right)\nonumber\\
%\\&=& \int_0^\alpha \ud q \left(q^2 \sqrt{q^2+\md^2} -q^3 - \md^2 q/2\right)\left( -\frac{q}{4} + \frac{q^3}{48}  -\frac{q^5}{480} + \dots\right) \nonumber
%\\
&=
\sum_{n=1}^\infty \frac{E_n(0)}{2n!} \int_0^\alpha \ud q q^{n} \left(q^2 \sqrt{q^2+\md^2} -q^3 - \md^2 q/2\right)\nonumber
\\&= \frac{\md^4}{32} \left[\alpha  - \frac{\alpha^3}{36} +\frac{\alpha^5}{600} \dots\right]  - \frac{\md^5}{30} + \dots\nonumber
\end{align}
\begin{eqnarray}
I_3(\md) &\equiv& \int_\alpha^\infty \ud q \frac{q^2 \sqrt{q^2+\md^2} - q^3 - \md^2q /2}{e^q+1}
\\&=& \sum_{n=2}^\infty \begm 1/2 \\ n \enm m^{2n}\int_\alpha^\infty \ud q \frac{q^{3-2n} }{e^q+1}.  \nonumber
\end{eqnarray}
The result is independent of $\alpha$, and evaluates numerically to
\begin{multline}
I(\md) = \frac{7}{120}\pi^4 + \frac{\pi^2}{24} \md^2 + \frac{\md^4\ln(\md)}{16} \\
- 0.0198446\md^4 - \frac{\md^5}{30} +
 0.0066606\md^6
 - \frac{\md^7}{630} + \dots.
\end{multline}
Thus the next term above the leading mass correction we consider in
the paper is ${\cal O}(\md^4\ln(\md))$. A similar approach can be
followed for a mass expansion of the pressure using
\begin{multline}
\int_0^\infty \ud q  \frac{q^4(q^2+\md^2)^{-1/2}}{e^q + 1} =
 \frac{7}{120}\pi^4 -   \frac{\pi^2}{24} \md^2
 - \frac{3\md^4\ln(\md)}{16} \\
-.00296608\md^4 + \frac{2\md^5}{15}  - 0.033303\md^6 + \frac{\md^7}{105} + \dots.
\end{multline}
\vfill

\FloatBarrier
\section{Initial Conditions}
\label{app:initial}

Here we present initial series solutions for the regular modes for
scalar, vector and tensor perturbations in cosmology. We allow for two
significantly different neutrino mass eigenstates, allowing us to
describe most of the possibilities of the neutrino mass hierarchy. The
solutions are correct to order $m^2$ in the neutrino mass. For space
we have only included the terms up to second order or the first
non-zero term up to order $\tau^3$.

We include the standard matter species which we generally denote with
subscripts: photons ($\gamma$), baryons ($b$), cold dark matter ($c$),
massless neutrinos ($\nu$) and massive neutrinos ($n$). Our solutions
are for after neutrino decoupling; we discuss the pre-decoupling
behaviour in the presence of magnetic fields in the
Section~\ref{sec:magneticmodes}. To solve the evolution of the
background equation we solve the Friedmann equations for the scale
factor. The solution to order $\bar{m}^2$ in the neutrino mass is
\begin{multline}
  a(\tau) = a_0\frac{\Omega_r}{\Omega_m} \biggl[\omega\tau +\frac{1}{4} \omega^2\tau^2 \\+ \frac{1}{12} R_n\frac{\Omega _r^2}{\Omega _m^2} \bar{m}^2 \omega^3\tau^3 + \frac{1}{96} R_n\frac{\Omega _r^2}{\Omega _m^2} \bar{m}^2 \omega^4\tau^4 \biggr]
\end{multline}
where we choose some time to fix the values of $\Omega_r =
\Omega_\gamma + \Omega_\nu + \Omega_n$ and $\Omega_m = \Omega_b +
\Omega_c$ and $R_n = \Omega_n / \Omega_r$. We have used the standard
definition of $\Omega_x = \rho_x /\rho_{\text{cr}}$ the ratio of the
density of species $x$ to the critical density. In the above we also
use the definition of $\omega = \Omega_m \clh_0 / \sqrt{\Omega_r}$.

To give our series solutions we will also use the definitions of
$R_\gamma = \Omega_\gamma / \Omega_r$, $R_\nu = \Omega_\nu /
\Omega_r$, $R_t = (\Omega_\nu + \Omega_n) / \Omega_r$, $R_c = \Omega_c
/ \Omega_m $ and $R_b = \Omega_b / \Omega_m$.

The solutions were calculated using Mathematica, and the notebooks and
relevent packages used can be found at \url{http://camb.info/jrs}.
  
\begin{widetext}  
\subsection{Scalar Initial Conditions}
There are six regular scalar modes, one adiabatic, four isocurvature,
and one magnetic. For comparison to other results we give our
solutions in the synchronous gauge \cite{Bucher:2000,Paoletti:2008}
with the standard potentials $h$ and $\eta$, commonly used for its
numerical robustness. This has the further advantage that the neutrino
velocity isocurvature mode is completely regular as $\tau \rightarrow
0$ \cite{Bucher:2000}. We also give the Bardeen potentials used in the
text $\Psi$ and $\Phi$.

\subsubsection*{Adiabatic Mode}
\begin{align*}
h(\tau)= & \frac{1}{2} k^2 \tau^2+O(\tau^3)\\ 
\eta (\tau)= & 1-\frac{5+4 R_t}{12 (15+4 R_t)} k^2 \tau^2+O(\tau^3)\\ 
\delta_c(\tau)= & -\frac{1}{4} k^2 \tau^2+O(\tau^3)\\ 
v_c(\tau)= & 0\\ 
\delta_n(\tau)= & -\frac{1}{3} k^2 \tau^2+O(\tau^3)\\ 
v_n(\tau)= & -\frac{23+4 R_t}{36 (15+4 R_t)} k^3 \tau^3+O(\tau^4)\\ 
\Pi_n(\tau)= & \frac{4}{15+4 R_t} k^2 \tau^2+O(\tau^3)\\ 
F_{n 3}(\tau)= & \frac{4}{3 (15+4 R_t)} k^3 \tau^3+O(\tau^4)\\ 
\delta_{\nu}(\tau)= & -\frac{1}{3} k^2 \tau^2+O(\tau^3)\\ 
v_{\nu}(\tau)= & -\frac{23+4 R_t}{36 (15+4 R_t)} k^3 \tau^3+O(\tau^4)\\ 
\Pi_{\nu}(\tau)= & \frac{4}{15+4 R_t} k^2 \tau^2+O(\tau^3)\\ 
F_{\nu 3}(\tau)= & \frac{4}{3 (15+4 R_t)} k^3 \tau^3+O(\tau^4)\\ 
\delta_b(\tau)= & -\frac{1}{4} k^2 \tau^2+O(\tau^3)\\ 
v_b(\tau)= & -\frac{1}{36} k^3 \tau^3+O(\tau^4)\\ 
\delta_{\gamma}(\tau)= & -\frac{1}{3} k^2 \tau^2+O(\tau^3)\\ 
v_{\gamma}(\tau)= & -\frac{1}{36} k^3 \tau^3+O(\tau^4)\\ 
\Psi (\tau)= & \frac{10}{15+4 R_t}+\frac{25 (-3+8 R_t)}{8 (15+2 R_t) (15+4 R_t)} \omega \tau+O(\tau^2)\\ 
\Phi (\tau)= & \frac{2 (5+2 R_t)}{15+4 R_t}-\frac{5 (15+16 R_t)}{8 (15+2 R_t) (15+4 R_t)} \omega \tau+O(\tau^2)
\end{align*}

\subsubsection*{CDM Isocurvature Mode}
\begin{align*}
h(\tau)= & R_c \omega \tau-\frac{3}{8} \left(R_c \omega ^2\right) \tau^2+O(\tau^3)\\ 
\eta (\tau)= & -\frac{1}{6} (R_c \omega ) \tau+\frac{1}{16} R_c \omega ^2 \tau^2+O(\tau^3)\\ 
\delta_c(\tau)= & 1-\frac{1}{2} (R_c \omega ) \tau+\frac{3}{16} R_c \omega ^2 \tau^2+O(\tau^3)\\ 
v_c(\tau)= & 0\\ 
\delta_n(\tau)= & -\frac{2}{3} (R_c \omega ) \tau+\frac{1}{4} R_c \omega ^2 \tau^2+O(\tau^3)\\ 
v_n(\tau)= & -\frac{1}{12} (k R_c \omega ) \tau^2+O(\tau^3)\\ 
\Pi_n(\tau)= & -\frac{k^2 R_c \omega \tau^3}{15+2 R_t}+O(\tau^4)\\ 
F_{n 3}(\tau)= & O(\tau^4)\\ 
\delta_{\nu}(\tau)= & -\frac{2}{3} (R_c \omega ) \tau+\frac{1}{4} R_c \omega ^2 \tau^2+O(\tau^3)\\ 
v_{\nu}(\tau)= & -\frac{1}{12} (k R_c \omega ) \tau^2+O(\tau^3)\\ 
\Pi_{\nu}(\tau)= & -\frac{k^2 R_c \omega \tau^3}{15+2 R_t}+O(\tau^4)\\ 
F_{\nu 3}(\tau)= & O(\tau^4)\\ 
\delta_b(\tau)= & -\frac{1}{2} (R_c \omega ) \tau+\frac{3}{16} R_c \omega ^2 \tau^2+O(\tau^3)\\ 
v_b(\tau)= & -\frac{1}{12} (k R_c \omega ) \tau^2+O(\tau^3)\\ 
\delta_{\gamma}(\tau)= & -\frac{2}{3} (R_c \omega ) \tau+\frac{1}{4} R_c \omega ^2 \tau^2+O(\tau^3)\\ 
v_{\gamma}(\tau)= & -\frac{1}{12} (k R_c \omega ) \tau^2+O(\tau^3)\\ 
\Psi (\tau)= & \frac{R_c (-15+4 R_t) \omega \tau}{8 (15+2 R_t)}+O(\tau^2)\\ 
\Phi (\tau)= & -\frac{(R_c (15+4 R_t) \omega ) \tau}{8 (15+2 R_t)}+O(\tau^2)
\end{align*}

\subsubsection*{Baryon Isocurvature Mode}
Baryon isocurvature modes are essentially observationally
indistinguishable from a rescaled CDM isocurvature mode. This is
because the compensated mode (with $\delta \rho_b = - \delta \rho_c$)
gives only a small contributions at small scales, primarily from the baryon
pressure and second order effects \cite{Gordon:2003}.

  \begin{align*}
h(\tau)= & R_b \omega \tau-\frac{3}{8} \left(R_b \omega ^2\right) \tau^2+O(\tau^3)\\ 
\eta (\tau)= & -\frac{1}{6} (R_b \omega ) \tau+\frac{1}{16} R_b \omega ^2 \tau^2+O(\tau^3)\\ 
\delta_c(\tau)= & -\frac{1}{2} (R_b \omega ) \tau+\frac{3}{16} R_b \omega ^2 \tau^2+O(\tau^3)\\ 
v_c(\tau)= & 0\\ 
\delta_n(\tau)= & -\frac{2}{3} (R_b \omega ) \tau+\frac{1}{4} R_b \omega ^2 \tau^2+O(\tau^3)\\ 
v_n(\tau)= & -\frac{1}{12} (k R_b \omega ) \tau^2+O(\tau^3)\\ 
\Pi_n(\tau)= & -\frac{k^2 R_b \omega \tau^3}{15+2 R_t}+O(\tau^4)\\ 
F_{n 3}(\tau)= & O(\tau^4)\\ 
\delta_{\nu}(\tau)= & -\frac{2}{3} (R_b \omega ) \tau+\frac{1}{4} R_b \omega ^2 \tau^2+O(\tau^3)\\ 
v_{\nu}(\tau)= & -\frac{1}{12} (k R_b \omega ) \tau^2+O(\tau^3)\\ 
\Pi_{\nu}(\tau)= & -\frac{k^2 R_b \omega \tau^3}{15+2 R_t}+O(\tau^4)\\ 
F_{\nu 3}(\tau)= & O(\tau^4)\\ 
\delta_b(\tau)= & 1-\frac{1}{2} (R_b \omega ) \tau+\frac{3}{16} R_b \omega ^2 \tau^2+O(\tau^3)\\ 
v_b(\tau)= & -\frac{1}{12} (k R_b \omega ) \tau^2+O(\tau^3)\\ 
\delta_{\gamma}(\tau)= & -\frac{2}{3} (R_b \omega ) \tau+\frac{1}{4} R_b \omega ^2 \tau^2+O(\tau^3)\\ 
v_{\gamma}(\tau)= & -\frac{1}{12} (k R_b \omega ) \tau^2+O(\tau^3)\\ 
\Psi (\tau)= & \frac{R_b (-15+4 R_t) \omega \tau}{8 (15+2 R_t)}+O(\tau^2)\\ 
\Phi (\tau)= & -\frac{(R_b (15+4 R_t) \omega ) \tau}{8 (15+2 R_t)}+O(\tau^2)
\end{align*}

\subsubsection*{Neutrino Isocurvature Mode}
\begin{align*}
h(\tau)= & \frac{3 R_n \omega ^2 \bar{m}^2 \Omega_r^2 \tau^2}{16 \Omega_m^2}+O(\tau^3)\\ 
\eta (\tau)= & \left(-\frac{k^2 R_t}{6 (15+4 R_t)}-\frac{R_n \omega ^2 \bar{m}^2 \Omega_r^2}{32 \Omega_m^2}\right) \tau^2+O(\tau^3)\\ 
\delta_c(\tau)= & -\frac{3 \left(R_n \omega ^2 \bar{m}^2 \Omega_r^2\right) \tau^2}{32 \Omega_m^2}+O(\tau^3)\\ 
v_c(\tau)= & 0\\ 
\delta_n(\tau)= & 1+\left(-\frac{k^2}{6}-\frac{(2+R_n) \omega ^2 \bar{m}^2 \Omega_r^2}{8 \Omega_m^2}\right) \tau^2+O(\tau^3)\\ 
v_n(\tau)= & \frac{k \tau}{4}+O(\tau^3)\\ 
\Pi_n(\tau)= & \frac{3 k^2 \tau^2}{15+4 R_t}+O(\tau^3)\\ 
F_{n 3}(\tau)= & \frac{k^3 \tau^3}{15+4 R_t}+O(\tau^4)\\ 
\delta_{\nu}(\tau)= & 1+\left(-\frac{k^2}{6}-\frac{R_n \omega ^2 \bar{m}^2 \Omega_r^2}{8 \Omega_m^2}\right) \tau^2+O(\tau^3)\\ 
v_{\nu}(\tau)= & \frac{k \tau}{4}+O(\tau^3)\\ 
\Pi_{\nu}(\tau)= & \frac{3 k^2 \tau^2}{15+4 R_t}+O(\tau^3)\\ 
F_{\nu 3}(\tau)= & \frac{k^3 \tau^3}{15+4 R_t}+O(\tau^4)\\ 
\delta_b(\tau)= & \left(\frac{k^2 R_t}{8 R_\gamma}-\frac{3 R_n \omega ^2 \bar{m}^2 \Omega_r^2}{32 \Omega_m^2}\right) \tau^2+O(\tau^3)\\ 
v_b(\tau)= & -\frac{(k R_t) \tau}{4 R_\gamma}+\frac{3 k R_b R_t \omega \tau^2}{16 R_\gamma^2}+O(\tau^3)\\ 
\delta_{\gamma}(\tau)= & -\frac{R_t}{R_\gamma}+\left(\frac{k^2 R_t}{6 R_\gamma}-\frac{R_n \omega ^2 \bar{m}^2 \Omega_r^2}{8 \Omega_m^2}\right) \tau^2+O(\tau^3)\\ 
v_{\gamma}(\tau)= & -\frac{(k R_t) \tau}{4 R_\gamma}+\frac{3 k R_b R_t \omega \tau^2}{16 R_\gamma^2}+O(\tau^3)\\ 
\Psi (\tau)= & -\frac{2 R_t}{15+4 R_t}+\frac{(75-2 R_t) R_t \omega \tau}{4 \left(225+90 R_t+8 R_t^2\right)}+O(\tau^2)\\ 
\Phi (\tau)= & \frac{R_t}{15+4 R_t}+\frac{R_t (-15+2 R_t) \omega \tau}{4 \left(225+90 R_t+8 R_t^2\right)}+O(\tau^2)
\end{align*}

\subsubsection*{Neutrino Velocity Isocurvature Mode}
Despite the apparent singularities in the potentials $\Psi$ and
$\Phi$, the mode is physical with a regular comoving curvature
perturbation. However, as the neutrinos are strongly coupled to the
photons prior to decoupling it is challenging to find a mechanism to
source this mode.

\begin{align*}
h(\tau)= & \frac{3 k R_b R_t \omega \tau^2}{8 R_\gamma}+O(\tau^3)\\ 
\eta (\tau)= & -\frac{4 (k R_t) \tau}{3 (5+4 R_t)}-\frac{\left(k R_t \left(75 R_b+80 R_b R_t+16 R_b R_t^2-80 R_\gamma\right) \omega \right) \tau^2}{16 ((5+4 R_t) (15+4 R_t) R_\gamma)}+O(\tau^3)\\ 
\delta_c(\tau)= & -\frac{3 (k R_b R_t \omega ) \tau^2}{16 R_\gamma}+O(\tau^3)\\ 
v_c(\tau)= & 0\\ 
\delta_n(\tau)= & -\frac{4 k \tau}{3}-\frac{(k R_b R_t \omega ) \tau^2}{4 R_\gamma}+O(\tau^3)\\ 
v_n(\tau)= & 1+\left(-\frac{k^2 (9+4 R_t)}{6 (5+4 R_t)}-\frac{\omega ^2 \bar{m}^2 \Omega_r^2}{4 \Omega_m^2}\right) \tau^2+O(\tau^3)\\ 
\Pi_n(\tau)= & \frac{8 k \tau}{5+4 R_t}+\frac{24 k R_t \omega \tau^2}{(5+4 R_t) (15+4 R_t)}+O(\tau^3)\\ 
F_{n 3}(\tau)= & \frac{4 k^2 \tau^2}{5+4 R_t}+O(\tau^3)\\ 
\delta_{\nu}(\tau)= & -\frac{4 k \tau}{3}-\frac{(k R_b R_t \omega ) \tau^2}{4 R_\gamma}+O(\tau^3)\\ 
v_{\nu}(\tau)= & 1-\frac{\left(k^2 (9+4 R_t)\right) \tau^2}{6 (5+4 R_t)}+O(\tau^3)\\ 
\Pi_{\nu}(\tau)= & \frac{8 k \tau}{5+4 R_t}+\frac{24 k R_t \omega \tau^2}{(5+4 R_t) (15+4 R_t)}+O(\tau^3)\\ 
F_{\nu 3}(\tau)= & \frac{4 k^2 \tau^2}{5+4 R_t}+O(\tau^3)\\ 
\delta_b(\tau)= & \frac{k R_t \tau}{R_\gamma}+\frac{3 k R_b (-3+R_t) R_t \omega \tau^2}{16 R_\gamma^2}+O(\tau^3)\\ 
v_b(\tau)= & -\frac{R_t}{R_\gamma}+\frac{3 R_b R_t \omega \tau}{4 R_\gamma^2}+\left(\frac{k^2 R_t}{6 R_\gamma}-\frac{3 R_b R_t (3 R_b-R_\gamma) \omega ^2}{16 R_\gamma^3}\right) \tau^2+O(\tau^3)\\ 
\delta_{\gamma}(\tau)= & \frac{4 k R_t \tau}{3 R_\gamma}+\frac{k R_b (-3+R_t) R_t \omega \tau^2}{4 R_\gamma^2}+O(\tau^3)\\ 
v_{\gamma}(\tau)= & -\frac{R_t}{R_\gamma}+\frac{3 R_b R_t \omega \tau}{4 R_\gamma^2}+\left(\frac{k^2 R_t}{6 R_\gamma}-\frac{3 R_b R_t (3 R_b-R_\gamma) \omega ^2}{16 R_\gamma^3}\right) \tau^2+O(\tau^3)\\ 
\Psi (\tau)= & -\frac{4 R_t}{(k (5+4 R_t)) \tau}-\frac{R_t (-45+4 R_t) \omega}{k (5+4 R_t) (15+4 R_t)} + O(\tau)\\ 
\Phi (\tau)= & \frac{4 R_t}{k (5+4 R_t) \tau}+\frac{R_t (-15+4 R_t) \omega}{k (5+4 R_t) (15+4 R_t)}+O(\tau)
\end{align*}

In theory we can define isocurvature modes in the neutrino anisotropic
stress and higher multipole moments, though with no reasonable
mechanism to produce them we will omit them.

\subsubsection*{Compensated Magnetic Modes}
For the compensated magnetic mode, we treat it like an isocurvature
mode with $\eta \rightarrow 0$ at very early times. For the density
sourced modes this gives
\begin{align*}
h(\tau)= & -\frac{3}{4} (R_\gamma \omega ) \tau+\left(\frac{9 R_\gamma \omega ^2}{32}-\frac{3 R_n R_\gamma \omega ^2 \bar{m}^2 \Omega_r^2}{16 \Omega_m^2}\right) \tau^2+O(\tau^3)\\ 
\eta (\tau)= & \frac{R_\gamma \omega \tau}{8}+\left(\frac{k^2 R_t R_\gamma}{6 (15+4 R_t)}-\frac{3 R_\gamma \omega ^2}{64}+\frac{R_n R_\gamma \omega ^2 \bar{m}^2 \Omega_r^2}{32 \Omega_m^2}\right) \tau^2+O(\tau^3)\\ 
\delta_c(\tau)= & -\frac{3 R_\gamma}{4}+\frac{3 R_\gamma \omega \tau}{8}+\left(-\frac{9 R_\gamma \omega ^2}{64}+\frac{3 R_n R_\gamma \omega ^2 \bar{m}^2 \Omega_r^2}{32 \Omega_m^2}\right) \tau^2+O(\tau^3)\\ 
v_c(\tau)= & 0\\ 
\delta_n(\tau)= & -R_\gamma+\frac{R_\gamma \omega \tau}{2}+\left(\frac{k^2 R_\gamma}{6}-\frac{3 R_\gamma \omega ^2}{16}+\frac{(2+R_n) R_\gamma \omega ^2 \bar{m}^2 \Omega_r^2}{8 \Omega_m^2}\right) \tau^2+O(\tau^3)\\ 
v_n(\tau)= & -\frac{1}{4} (k R_\gamma) \tau+\frac{1}{16} k R_\gamma \omega \tau^2+O(\tau^3)\\ 
\Pi_n(\tau)= & -\frac{3 \left(k^2 R_\gamma\right) \tau^2}{15+4 R_t}+O(\tau^3)\\ 
F_{n 3}(\tau)= & -\frac{k^3 R_\gamma \tau^3}{15+4 R_t}+O(\tau^4)\\ 
\delta_{\nu}(\tau)= & -R_\gamma+\frac{R_\gamma \omega \tau}{2}+\left(\frac{k^2 R_\gamma}{6}-\frac{3 R_\gamma \omega ^2}{16}+\frac{R_n R_\gamma \omega ^2 \bar{m}^2 \Omega_r^2}{8 \Omega_m^2}\right) \tau^2+O(\tau^3)\\ 
v_{\nu}(\tau)= & -\frac{1}{4} (k R_\gamma) \tau+\frac{1}{16} k R_\gamma \omega \tau^2+O(\tau^3)\\ 
\Pi_{\nu}(\tau)= & -\frac{3 \left(k^2 R_\gamma\right) \tau^2}{15+4 R_t}+O(\tau^3)\\ 
F_{\nu 3}(\tau)= & -\frac{k^3 R_\gamma \tau^3}{15+4 R_t}+O(\tau^4)\\ 
\delta_b(\tau)= & -\frac{3 R_\gamma}{4}+\frac{3 R_\gamma \omega \tau}{8}+\left(-\frac{k^2 R_t}{8}-\frac{9 R_\gamma \omega ^2}{64}+\frac{3 R_n R_\gamma \omega ^2 \bar{m}^2 \Omega_r^2}{32 \Omega_m^2}\right) \tau^2+O(\tau^3)\\ 
v_b(\tau)= & \frac{k R_t \tau}{4}+\frac{k \left(1-2 R_t-3 R_b R_t+R_t^2\right) \omega \tau^2}{16 R_\gamma}+O(\tau^3)\\ 
\delta_{\gamma}(\tau)= & -R_\gamma+\frac{R_\gamma \omega \tau}{2}+\left(-\frac{k^2 R_t}{6}-\frac{3 R_\gamma \omega ^2}{16}+\frac{R_n R_\gamma \omega ^2 \bar{m}^2 \Omega_r^2}{8 \Omega_m^2}\right) \tau^2+O(\tau^3)\\ 
v_{\gamma}(\tau)= & \frac{k R_t \tau}{4}+\frac{k \left(1-2 R_t-3 R_b R_t+R_t^2\right) \omega \tau^2}{16 R_\gamma}+O(\tau^3)\\ 
\Psi (\tau)= & \frac{2 R_t R_\gamma}{15+4 R_t}+\frac{(675-8 R_t (75+4 R_t)) R_\gamma \omega \tau}{32 (15+2 R_t) (15+4 R_t)}+O(\tau^2)\\ 
\Phi (\tau)= & -\frac{R_t R_\gamma}{15+4 R_t}+\frac{(675+32 R_t (15+R_t)) R_\gamma \omega \tau}{32 (15+2 R_t) (15+4 R_t)}+O(\tau^2)
\end{align*}

The anisotropic stress sourced modes are
\begin{align*}
h(\tau)= & \frac{1}{60} k^2 R_b \omega \tau^3+O(\tau^4)\\ 
\eta (\tau)= & \left(-\frac{55 k^2 R_\gamma}{252 (15+4 R_t)}-\frac{5 R_n R_\gamma \omega ^2 \bar{m}^2 \Omega_r^2}{24 R_t (15+4 R_t) \Omega_m^2}\right) \tau^2+O(\tau^3)\\ 
\delta_c(\tau)= & -\frac{1}{120} \left(k^2 R_b \omega \right) \tau^3+O(\tau^4)\\ 
v_c(\tau)= & 0\\ 
\delta_n(\tau)= & -\frac{\left(k^2 R_\gamma\right) \tau^2}{9 R_t}+O(\tau^3)\\ 
v_n(\tau)= & \frac{k R_\gamma \tau}{6 R_t}+O(\tau^3)\\ 
\Pi_n(\tau)= & -\frac{R_\gamma}{R_t}+\left(\frac{55 k^2 R_\gamma}{14 R_t (15+4 R_t)}-\frac{(15+4 R_n+4 R_t) R_\gamma \omega ^2 \bar{m}^2 \Omega_r^2}{4 R_t (15+4 R_t) \Omega_m^2}\right) \tau^2+O(\tau^3)\\ 
F_{n 3}(\tau)= & -\frac{k R_\gamma \tau}{R_t}+O(\tau^3)\\ 
\delta_{\nu}(\tau)= & -\frac{\left(k^2 R_\gamma\right) \tau^2}{9 R_t}+O(\tau^3)\\ 
v_{\nu}(\tau)= & \frac{k R_\gamma \tau}{6 R_t}+O(\tau^3)\\ 
\Pi_{\nu}(\tau)= & -\frac{R_\gamma}{R_t}+\left(\frac{55 k^2 R_\gamma}{14 R_t (15+4 R_t)}-\frac{R_n R_\gamma \omega ^2 \bar{m}^2 \Omega_r^2}{R_t (15+4 R_t) \Omega_m^2}\right) \tau^2+O(\tau^3)\\ 
F_{\nu 3}(\tau)= & -\frac{k R_\gamma \tau}{R_t}+O(\tau^3)\\ 
\delta_b(\tau)= & \frac{k^2 \tau^2}{12}+O(\tau^3)\\ 
v_b(\tau)= & -\frac{k \tau}{6}+\frac{k R_b \omega \tau^2}{8 R_\gamma}+O(\tau^3)\\ 
\delta_{\gamma}(\tau)= & \frac{k^2 \tau^2}{9}+O(\tau^3)\\ 
v_{\gamma}(\tau)= & -\frac{k \tau}{6}+\frac{k R_b \omega \tau^2}{8 R_\gamma}+O(\tau^3)\\ 
\Psi (\tau)= & \left(-\frac{55 R_\gamma}{21 (15+4 R_t)}-\frac{5 R_n R_\gamma \omega ^2 \bar{m}^2 \Omega_r^2}{2 k^2 R_t (15+4 R_t) \Omega_m^2}\right) \\ &+\left(-\frac{55 (-75+2 R_t) R_\gamma \omega}{168 (15+2 R_t) (15+4 R_t)}-\frac{25 R_n (-3+8 R_t) R_\gamma \omega ^3 \bar{m}^2 \Omega_r^2}{32 k^2 R_t (15+2 R_t) (15+4 R_t) \Omega_m^2}\right) \tau+O(\tau^2)\\ 
\Phi (\tau)= & \left(\frac{55 R_\gamma}{42 (15+4 R_t)}+\frac{5 R_n R_\gamma \omega ^2 \bar{m}^2 \Omega_r^2}{4 k^2 R_t (15+4 R_t) \Omega_m^2}\right) \\ &+\left(\frac{55 (-15+2 R_t) R_\gamma \omega}{168 (15+2 R_t) (15+4 R_t)}+\frac{5 R_n (15+16 R_t) R_\gamma \omega ^3 \bar{m}^2 \Omega_r^2}{32 k^2 R_t (15+2 R_t) (15+4 R_t) \Omega_m^2}\right) \tau+O(\tau^2)
\end{align*}

\subsection{Vector Initial Conditions}

There are two regular vector modes, a vorticity mode which is the
vector equivalent of the neutrino velocity isocurvature mode, and a
magnetic mode compensating the magnetic anisotropic stress
$\Pi_B^\brsc{1}$. We give the solutions in terms of the gauge
invariant variables used earlier.

\subsubsection*{Vorticity Mode}
For the same reasons as the neutrino velocity mode, the existence of
this type of perturbation is highly unlikely.
\begin{align*}
\sigma^\brsc{1}(\tau)= & 1-\frac{15 \omega \tau}{30+8 R_t}+\left(-\frac{15 k^2}{420+56 R_t}+\frac{(675-60 R_t) \omega ^2}{16 \left(225+90 R_t+8 R_t^2\right)}-\frac{R_n (5+6 R_t) \omega ^2 \bar{m}^2 \Omega_r^2}{2 R_t (15+2 R_t) \Omega_m^2}\right) \tau^2+O(\tau^3)\\ 
\Omega^\brsc{1}_{c}(\tau)= & 0\\ 
\Omega^\brsc{1}_{n}(\tau)= & \left(1+\frac{5}{4 R_t}\right)+\left(-\frac{k^2}{8 R_t}-\frac{(5+4 R_t) \omega ^2 \bar{m}^2 \Omega_r^2}{16 R_t \Omega_m^2}\right) \tau^2+O(\tau^3)\\ 
\Pi^\brsc{1}_{n}(\tau)= & \frac{2 k \tau}{R_t}+\frac{6 k \omega \tau^2}{15+4 R_t}+O(\tau^3)\\ 
F_{n 3}(\tau)= & \frac{\sqrt{\frac{2}{3}} k^2 \tau^2}{R_t}+O(\tau^3)\\ 
\Omega^\brsc{1}_{\nu}(\tau)= & \left(1+\frac{5}{4 R_t}\right)-\frac{k^2 \tau^2}{8 R_t}+O(\tau^3)\\ 
\Pi^\brsc{1}_{\nu}(\tau)= & \frac{2 k \tau}{R_t}+\frac{6 k \omega \tau^2}{15+4 R_t}+O(\tau^3)\\ 
F_{\nu 3}(\tau)= & \frac{\sqrt{\frac{2}{3}} k^2 \tau^2}{R_t}+O(\tau^3)\\ 
\Omega^\brsc{1}_{b}(\tau)= & -\frac{5+4 R_t}{4 R_\gamma}+\frac{3 R_b (5+4 R_t) \omega \tau}{16 R_\gamma^2}-\frac{3 \left(R_b (5+4 R_t) (3 R_b-R_\gamma) \omega ^2\right) \tau^2}{64 R_\gamma^3}+O(\tau^3)\\ 
\Omega^\brsc{1}_{\gamma}(\tau)= & -\frac{5+4 R_t}{4 R_\gamma}+\frac{3 R_b (5+4 R_t) \omega \tau}{16 R_\gamma^2}-\frac{3 \left(R_b (5+4 R_t) (3 R_b-R_\gamma) \omega ^2\right) \tau^2}{64 R_\gamma^3}+O(\tau^3)
\end{align*}

\subsubsection*{Compensated Magnetic Mode}
\begin{align*}
\sigma^\brsc{1}(\tau)= & \left(\frac{15 k R_\gamma}{210+56 R_t}+\frac{5 R_n R_\gamma \omega ^2 \bar{m}^2 \Omega_r^2}{4 k R_t (15+4 R_t) \Omega_m^2}\right) \tau \\ &+\left(-\frac{225 k R_\gamma \omega}{28 \left(225+90 R_t+8 R_t^2\right)}+\frac{15 R_n (-5+4 R_t) R_\gamma \omega ^3 \bar{m}^2 \Omega_r^2}{32 k R_t \left(225+90 R_t+8 R_t^2\right) \Omega_m^2}\right) \tau^2+O(\tau^3)\\ 
\Omega^\brsc{1}_{c}(\tau)= & 0\\ 
\Omega^\brsc{1}_{n}(\tau)= & \frac{k R_\gamma \tau}{8 R_t}+O(\tau^3)\\ 
\Pi^\brsc{1}_{n}(\tau)= & -\frac{R_\gamma}{R_t}+\left(\frac{45 k^2 R_\gamma}{210 R_t+56 R_t^2}-\frac{(15+4 R_n+4 R_t) R_\gamma \omega ^2 \bar{m}^2 \Omega_r^2}{4 R_t (15+4 R_t) \Omega_m^2}\right) \tau^2+O(\tau^3)\\ 
F_{n 3}(\tau)= & -\frac{\sqrt{\frac{2}{3}} k R_\gamma \tau}{R_t}+O(\tau^3)\\ 
\Omega^\brsc{1}_{\nu}(\tau)= & \frac{k R_\gamma \tau}{8 R_t}+O(\tau^3)\\ 
\Pi^\brsc{1}_{\nu}(\tau)= & -\frac{R_\gamma}{R_t}+\left(\frac{45 k^2 R_\gamma}{210 R_t+56 R_t^2}-\frac{R_n R_\gamma \omega ^2 \bar{m}^2 \Omega_r^2}{R_t (15+4 R_t) \Omega_m^2}\right) \tau^2+O(\tau^3)\\ 
F_{\nu 3}(\tau)= & -\frac{\sqrt{\frac{2}{3}} k R_\gamma \tau}{R_t}+O(\tau^3)\\ 
\Omega^\brsc{1}_{b}(\tau)= & -\frac{k \tau}{8}+\frac{3 k R_b \omega \tau^2}{32 R_\gamma}+O(\tau^3)\\ 
\Omega^\brsc{1}_{\gamma}(\tau)= & -\frac{k \tau}{8}+\frac{3 k R_b \omega \tau^2}{32 R_\gamma}+O(\tau^3)
\end{align*}

\subsection{Tensor Initial Conditions}

Only the photons and neutrinos can support tensor perturbations to
their energy momentum tensors and at times long before recombination
the photon anisotropic stress is negligible. Thus the species
affecting the tensor evolution are the neutrinos, and the magnetic
fields. This leaves us with one standard tensor mode, the
gravitational wave mode, and a compensated magnetic mode.

\subsubsection*{Gravitational Wave Mode}
\begin{align*}
H^\brsc{2}(\tau)= & 1-\frac{5}{2 (15+4 R_t)} k^2 \tau^2+O(\tau^3)\\ 
\Pi^\brsc{2}_{\nu}(\tau)= & \frac{4}{15+4 R_t} k^2 \tau^2+O(\tau^3)\\ 
F_{\nu 3}(\tau)= & \frac{2 \sqrt{5}}{3 (15+4 R_t)} k^3 \tau^3+O(\tau^4)\\ 
\Pi^\brsc{2}_{n}(\tau)= & \frac{4}{15+4 R_t} k^2 \tau^2+O(\tau^3)\\ 
F_{n 3}(\tau)= & \frac{2 \sqrt{5}}{3 (15+4 R_t)} k^3 \tau^3+O(\tau^4)
\end{align*}

\subsubsection*{Compensated Magnetic Mode}
\begin{align*}
H^\brsc{2}(\tau)= & \left(\frac{5 k^2 R_\gamma}{28 (15+4 R_t)}+\frac{5 R_n R_\gamma \omega ^2 \bar{m}^2 \Omega_r^2}{8 R_t (15+4 R_t) \Omega_m^2}\right) \tau^2+O(\tau^3)\\ 
\Pi^\brsc{2}_{\nu}(\tau)= & -\frac{R_\gamma}{R_t}+\left(\frac{15 k^2 R_\gamma}{210 R_t+56 R_t^2}-\frac{R_n R_\gamma \omega ^2 \bar{m}^2 \Omega_r^2}{R_t (15+4 R_t) \Omega_m^2}\right) \tau^2+O(\tau^3)\\ 
F_{\nu 3}(\tau)= & -\frac{\left(\sqrt{5} k R_\gamma\right) \tau}{2 R_t}+O(\tau^3)\\ 
\Pi^\brsc{2}_{n}(\tau)= & -\frac{R_\gamma}{R_t}+\left(\frac{15 k^2 R_\gamma}{210 R_t+56 R_t^2}-\frac{(15+4 R_n+4 R_t) R_\gamma \omega ^2 \bar{m}^2 \Omega_r^2}{4 R_t (15+4 R_t) \Omega_m^2}\right) \tau^2+O(\tau^3)\\ 
F_{n 3}(\tau)= & -\frac{\left(\sqrt{5} k R_\gamma\right) \tau}{2 R_t}+O(\tau^3)
\end{align*}

\end{widetext}

\bibliographystyle{arxiv}
\bibliography{neutrino,antony}

\end{document}